\pdfoutput=1
\documentclass[modern]{aastex61}

\received{\today}
\revised{\today}
\accepted{\today}

\submitjournal{ApJ}

\newcommand{\beq}{\begin{equation}}
\newcommand{\eeq}{\end{equation}}
\newcommand{\Mdot}{\dot{M}}

\newcommand{\kms}{\mbox{km s$^{-1}$}}

\newcommand{\Mo}{\mbox{M$_{\odot}$}}
\newcommand{\Moy}{\mbox{M$_{\odot}$ yr$^{-1}$}}

\shorttitle{Common Envelope Shaping of Planetary Nebulae. III.}

\shortauthors{Garc\'{\i}a-Segura et al.}

\begin{document}

\title{Common Envelope Shaping of Planetary Nebulae. III.

The Launching of  Jets in Proto-Planetary Nebulae}

\correspondingauthor{Guillermo Garc\'{\i}a-Segura}
\email{ggs@astrosen.unam.mx}

\author{Guillermo Garc\'{\i}a-Segura}
\affiliation{Instituto de Astronom\'{\i}a, Universidad Nacional Aut\'onoma
de M\'exico, Km. 107 Carr. Tijuana-Ensenada, 22860, Ensenada, B. C., Mexico}

\author{Ronald E. Taam}
\affiliation{Center for Interdisciplinary Exploration and Research in Astrophysics (CIERA), 
Department of Physics and Astronomy, 
Northwestern University, 2145 Sheridan Road, Evanston, IL 60208, USA}

\author{Paul M. Ricker}
\affiliation{Department of Astronomy, University of Illinois, 1002 W. Green St., Urbana, IL 61801, USA}

\begin{abstract}

We compute successfully the launching of two
magnetic winds from two circumbinary 
disks formed after a common envelope event. The launching is produced by the increase of magnetic pressure due to the collapse of the disks. 
The collapse is due to internal torques produced by
a weak poloidal magnetic field.
The first wind can be described as a wide jet, 
with  an average  mass-loss rate of 
$\sim 1.3 \times 10^{-7}$ \Moy\ and a maximum radial velocity of $\sim 230$ \kms. 
The outflow has a half-opening angle of $\sim 20^{\circ}$.
Narrow jets are also formed intermittently with velocities up to 3,000 \kms, 
with mass-loss rates of $\sim 6 \times 10^{-12} $ \Moy\ during short periods of time. 
The second wind can be described as a wide X-wind, 
with an 
average mass-loss rate of $\sim 1.68 \times 10^{-7}$ \Moy\ and a velocity 
of $\sim 30$ \kms. A narrow jet is also formed with 
a velocity of 250 \kms, and a mass-loss rates of $\sim 10^{-12} $ \Moy.

The computed  jets are  used to provide inflow boundary conditions for simulations of proto-planetary nebulae. The wide jet evolves into a molecular collimated outflow within a few astronomical units,
producing proto-planetary nebulae with bipolar, elongated shapes, whose kinetic energies reach  
$\sim 4 \times 10^{45}$ erg at 1,000 years.
Similarities with observed features in 
W43A, OH231.8+4.2, and Hen 3-1475 are discussed. 

The computed wide X-wind produces proto-planetary nebulae with slower expansion 
velocities,  with bipolar and elliptical shapes, and possible starfish type and quadrupolar morphology.

\end{abstract}


\section{Introduction} 
\label{sec:intro}

Since the early evidence for and detection of jets in planetary nebulae (PNe) (Feibelman 1985; Gieseking et al.\ 1985; Miranda \& Solf 1990; L\'opez et al.\ 1993), a number of studies have expanded upon the idea that jets in PNe and proto-planetary nebulae 
(PPNe) are common instead of something exotic, especially after observations made with the Hubble Space
Telescope (Borkowski et al.\ 1997; Sahai \& Trauger 1998)
and ground-based CO millimeter observations (Alcolea et al.\ 2000; 
Bujarraval et al.\ 2001), which more recently include the Atacama Large Millimeter Array (ALMA) (S\'anchez-Contreras et al.\ 2018; Tafoya et al.\ 2020).
The list of studies is large, and a recent compilation is 
given in Guerrero et al.\ (2020).

At the same time, early theoretical work on jets in PNe started to give the first explanations for this phenomenon
(Soker 1992; Pascoli 1993; Soker \& Livio, 1994; R\'o\.zyczka \& Franco 1996; Garc\'{\i}a-Segura 1997;
Lee \& Sahai 2003). 
However, despite the large number of observations about jets in PNe and PPNe, there is no theoretical work on the launching of jets in these scenarios. All past numerical experiments with PNe and PPNe  assumed an 
injection of momentum or energy at an inner boundary (Garc\'{\i}a-Segura et al.\ 2005; Garc\'{\i}a-Segura et al.\ 2018, paper I; 
Balick et al.\ 2020; Garc\'{\i}a-Segura et al.\ 2020, paper II, and references therein), 
and the launching process was not studied.

In paper II, we followed the evolution of a magnetized, ejected envelope after a common envelope (CE) event (Ricker \& Taam 2012), and we showed that the $\sim 10^4$ gauss magnetic field compressed into the circumbinary disk was promising for the launching of magnetic winds. Following that insight, we now build on that result (i.e., when a  circumbinary disk with an important 
toroidal magnetic field has been formed) and include the necessary and sufficient physical inputs to launch a magnetic wind (i.e., a weak poloidal magnetic 
field which is hypothesized to form from the toroidal field itself), as we will discuss in \S~3.

This article is a continuation of paper II and is structured as follows.  The
numerical scheme is described in \S~2 . The physical assumptions 
underlying the initial conditions are described in \S~3. 
The results of the numerical simulations are presented in \S~4.   
Finally, we discuss the numerical results, applying them to several observed PPNe, in \S~5 and provide the main conclusions in the last section.

\section{Numerical simulations}

The simulations have been performed using the magnetohydrodynamic code
ZEUS-3D (version 3.4), developed by M. L. Norman and the Laboratory for
Computational Astrophysics (Stone \& Norman 1992; Clarke 1996), with the same scheme used in paper I and paper II.
The simulations are run in a single processor mode on a Supermicro server.
Photoionization is not included in the present paper.

The computational grid for Models AF (``F'' for fast) and AS  (``S'' for slow) consists of $200 \times 200$ equally spaced zones in spherical coordinates 
$r$ and $\theta$ respectively, with an angular extent of $90^{\circ}$ and a
radial extent of $1.9 \times 10^{13}$ cm (1.27 AU). 
The radial resolution is $9.452 \times 10^{10} $ cm.
The inner boundary of the grid in Model AF lies at $ r= 9.5 \times10^{10}$ cm (0.00635 AU). 
Model AF  is aimed at
binaries with a short period or systems that lead to mergers. 
The inner boundary of the grid in Model AS lies at
$r = 6 \times 10^{11}$ cm (0.04 AU). 
Model AS  is aimed at the outcome of
the simulation by Ricker \& Taam (2012), where the binary has
an orbit of $\sim 0.02$ AU. 
We emphasize that the inner volume where the binary system produces a gap
between the stars and the circumbinary disk cannot be modeled in 2D, since it is not
axisymmetric, and for that reason we have located the inner boundaries of Models AF and AS at the inner edge of the circumbinary disk, to avoid including the gap in the computation.

Models AF and AS have a fixed grid during the evolution with outflow conditions on the inner and the outer boundaries.
This is done to allow accretion at the inner boundary and outflows at the outer boundary.

The computational grids for Models B$x$, C$x$, D$x$, E$x$, and F$x$ each have $800 \times 200$
initially equally spaced zones,  with an angular extent of $90^{\circ}$ and an
initial radial extent of $1.52 \times 10^{16}$ cm (1013 AU). 
These models have self-expanding grids with inflow conditions on the inner and  outer boundaries. These models' setups are original of this paper, but  include an interpolation of the ejected envelope from paper II (see below). 

At the inner boundary, which is located at $r=1.9\times10^{13}$~cm (1.27 AU), the 
time-averaged winds computed in  Models AF and AS  are injected. 
At the outer boundary of the moving grid the pre-existing AGB wind flows inward due to the more rapid expansion 
of the grid (typically of the order of 200 km~s$^{-1}$).
Runs in the B$x$, C$x$, E$x$, and F$x$ sequences have mass-loss rates of $10^{-6}$, $10^{-7}$, $10^{-8}$ and $10^{-9}$ \Moy, with the number $x$ of each model corresponding to the exponent of the mass-loss rate. The AGB wind velocity is assumed to be 10~km~s$^{-1}$. Model D9
uses $10^{-9}$ \Moy\ for a  pre-existing red giant (RG) wind 
with speed 10~km~s$^{-1}$.

Table 1 displays a summary of the models. We describe the initialization of the interior of each model grid in the next section.

\section{Physical assumptions underlying the initial conditions}

The initial conditions for Model AF are taken from 
Model A in paper II, 
at a time corresponding to 156.7 days after the
CE (or 100 days after the initial conditions for paper II, i.e., 56.7 days after CE). 
The initial conditions are displayed in Figure 1, when the circumbinary disk is already formed. Various physical 
quantities are displayed in Table 2. 
The initial conditions for Model AS are taken from a new model, with the inner boundary
located at 0.04 AU, but otherwise similar to Model A in paper II.
The central binary has at this stage a period of $\sim 5$ days and an orbital  
separation of $\sim 0.02 $ AU. The binary is not included in our grid, but the combined gravitational force 
is included as a point source at the center.
The disk is dominated by thermal pressure,
which is $\sim 10^2$ larger than the magnetic pressure.

Note that the 2-dimensional, axisymmetric configuration displayed in Figure 1 is in Keplerian rotation, and it is
stable while running the numerical code for several binary orbits. Since there is no friction,
there is no loss of angular momentum.  Therefore, no wind/jet or outflow will be launched with this setup, as we already computed in paper II.

In the initial conditions for Models AF and AS we introduce artificially a weak poloidal magnetic field equal to 0.5\%   of 
the toroidal field.\footnote{For clarity, if the initial magnetic field has spherical polar components $(0,0,B_\phi)$, we perturb it to $(0.005B_\phi\cos\theta, 0.005B_\phi\sin\theta, B_\phi)$.} The poloidal field will introduce internal torques that will be responsible for
the launching of the magnetic winds (see \S~4).
Note that in a  full 3-dimensional  computation, as already was discussed by Stone \& Norman (1994), 
nonaxisymmetric modes of the Parker instability may develop, causing toroidal flux loops to buoyantly rise out of the plane of the disk along the $\phi$ direction, producing a poloidal field from the 
toroidal one (similar to Figure 5 in Blandford \& Payne 1982 along the $\phi$ direction). Thus, in three dimensions the poloidal field would arise naturally, and this is the field that we introduce in our axisymmetric calculation.
Either a perpendicular poloidal field with respect to the disk or a dipolar poloidal field can be used in the 
calculation (Stone \& Norman 1994), since in both configurations the dominant  component of the poloidal 
field is $B_\theta$ near the equator ($\theta = 90^{ \circ} $).

It is worthwhile to mention that we do not inject any inflow or magnetic energy at the inner boundary 
of Models AF and AS, and the launching of the wind/jet is 
the result of the physical conditions and the gravitational force.

The initial conditions for Models B$x$, C$x$, D9, E$x$, and F$x$ are taken 
from Model C in paper II 
at a time when the ejected envelope has evolved to 176.7 days after the
CE (or 120 days after the initial conditions in paper II). 
At the inner boundary, we now inject the terminal wind computed in Models AF and
AS, which will be discussed in 
the next section. 

Since the grids are self-expanding in Models B$x$, C$x$, D9, E$x$, and F$x$, we have to assume a behavior for the winds at the inner 
boundary as a function of the radial distance, since either the outer or the inner boundaries are expanding. 
Models B$x$, D9, and E$x$ (option a) assume that the density in the wind drops as $1/r^2$ due to spherical divergence, and $B_\phi$ as  $1/r$
(Chevalier \& Luo 1994). In models C$x$ and F$x$ (option b), on the other hand, both $B_\phi$ and the density decline as $1/r^2$. 
This uncertainty reflects the fact that turbulence as well as kink and sausage 
instabilities  break the structure of the collimated jets (see for example a 3-dimensional calculation by
Huarte-Espinosa et al.\ 2012), reducing the efficiency of the toroidal magnetic field in collimating or accelerating the wind.

Finally, following the same reasoning as in paper I, Models B$x$, C$x$, E$x$, and
F$x$ consider CE scenarios where
an AGB is the primary star, so the initial expansion velocity of the ejected envelope has been reduced 
by a factor 2 to account for a lower escape velocity, while model D9 is 
for a RG star, with a much 
higher value  of  the escape velocity, as in the original calculation by Ricker \& Taam (2012).

\section{Results}

\subsection{Model AF. The launching of the wide jet}

At the beginning of Model AF (Figure 1), the disk is dominated by the thermal pressure,
i.e., the plasma  $\beta = P_{\rm ther}/P_{\rm mag} $  
$\sim 10^2$, and the disk is in  Keplerian rotation. Some physical quantities are presented in Table 2.
The weak poloidal magnetic field introduces internal torques in the disk, causing the loss of angular momentum.  Due to this magnetic
braking, the disk starts to collapse, increasing both its rotational velocity and its toroidal magnetic field due to its compression. 
In day 1 and 2, the plasma $\beta$ approaches 1 near the origin, while in day 3,  $\beta$  reaches values of 0.1  
at some points close to the origin, where the flow is clearly dominated by the pressure of the toroidal field. 
However, already in day 2 some launching is produced by a small
magnetocentrifugal effect. The magnetocentrifugal launching here is not very intense, but it is sufficient
to trigger the launching of the gas by the pressure of the toroidal magnetic field after day 3. 
It is interesting to note that this first magnetocentrifugal push occurs where the gas 
is nearly levitated from the disk by the toroidal magnetic pressure, and this occurs at the border of the inner disk, where the density drops sharply.

Figure 2 shows several physical quantities at  20 days from the start, by which time a well-established
magnetic wind has developed. The maximum acceleration of the launched wind 
occurs at $\sim 20^{\circ}$ from the polar axis. 
We note that a centrifugal barrier can 
be observed in the wind, since the gas never gets close to the polar axis.
The low density gas at the polar axis represents gas that is still collapsing from the original ejected envelope.

The third panel shows the ratio of the Alfv\'en velocity to the gas velocity.
It can be seen that the wind zones where the gas is sub-Alfv\'enic
(green-yellow-red) are the places where the launching 
takes place.
In these zones, the magnetic pressure dominates  the dynamics. 
The launching is clearly seen also in the fourth panel, where the radial velocity is displayed. In the co-latitude 
velocity (fifth panel), positive corresponds to motion 
towards the equator, while negative corresponds to motion towards the polar direction.

Figure 3 shows the developed wind at 44 days. The wind can be 
described as a wide jet, with a maximum radial velocity of 229 \kms\ (Table 3), and a mass-loss rate of $1.23 \times 10^{-7}$  \Moy\ on average. The outflow has a half-opening angle of $\sim 20^{\circ}$.

The wind is not steady.  In particular, starting at day 27, the wind develops  pulses, with small bursts, as 
shown in the time sequence of  Figure 4. 
A total of four pulses are computed, which is reflected in the time-dependent mass-loss rate displayed
in Figure 5. As seen in Figure 4, the second pulse occurs
at 1.6 days from the first, the third pulse 3 days later, and the fourth pulse occurs 4 days later than the third.

The pulses are caused as a consequence of the  collision of the gas with the polar axis (the collision that produce a narrow jet as described below),
producing a wave as a back reaction that travels back and forth in the $\theta$ direction.

In each collision, the upper part of the gas (the gas at larger distance from the center) is ejected by the compressed magnetic
pressure, and the lower part of the gas is accreted into the inner boundary. Note that these ejections
are not well collimated, since the topology of the magnetic field is not appropriate, i.e., the
toroidal field does not completely surround each ejection, in comparison with the narrow jet produced in the polar axis (see below).

At the inner boundary, the accreted flow reaches a mass accretion rate
of $4.35 \times 10^{-5}$ \Moy\  (Figure 5). Note that the accreted gas can flow 
toward the central binary, forming
one or two independent accretion disks around each star.

Figure 6  shows several physical quantities of the terminal wind at the outer
boundary. These are the values that we used in the simulations of PPNe described in \S~4.3.

\subsection{Model AF. The launching of narrow jets}

During the pulses described above, the gas that arrives directly to the 
polar axis experiences a larger hoop stress, as well as a higher compression, producing
a large magnetic pressure. As a consequence,
part of the gas is launched along the polar axis, and part of the gas in accreted along the pole. Here the topology of the toroidal magnetic field is able to collimate the gas very efficiently,
 since the gas is completely surrounded by the magnetic field aligned with the polar axis.
Under such
conditions, the gas is launched very rapidly, forming a narrow jet. 
These are the conditions for the ``plasma gun'' scenario (Contopoulos 1995).
The first narrow jet occurs after day 27 in the simulation. In total, 3 narrow jets are formed in the computation.

Figure 7 shows several physical quantities (gas density, toroidal magnetic field, radial velocity, and rotational velocity) at days 
27, 28, 29 and 30 (183.7, 184.7, 185.7, and 186.7 days after the CE event).
The gas at the inner disk in the equatorial region is rotating at 311 \kms, but the maximum rotational velocity
 in each snapshot reaches 500, 1,964, 1,148, and 1,248 \kms\ due to the high compression along the polar axis. The gas is then launched 
at very high speed by the excess of magnetic pressure in the polar direction, reaching 2,276 \kms\ in the third panel (day 29). The magnetic field snapshots reveal the increase of magnetic pressure close to the polar axis very clearly (days 28, 29, and 30), while
the density snapshots show the formation of a narrow jet (days 29 and 30). 
The mass-loss rate due to the narrow jet is very small, with 
a three-day time average of $6 \times 10^{-12}$ \Moy\ over the part of the outer boundary lying within ten angular zones about the polar axis.

In order to study further the formation of this jet, an additional simulation
starting at day 24 was carried out in which the grid was allowed to self-expand to track the bow shock of the jet.
The result is shown in Figure 8, just two days later than Figure 7. The maximum velocity here is 2,929 \kms. 
After this day, the jet  decelerates. It is clear that the speed here is much faster than the escape velocity of the 
binary, and it is 
an order of magnitude larger than the velocity of the wide jet.
However, the amount of mass in the jet is small compared to that of the wide jet. For this reason, 
we have used only the wide jet shown in Figure 6, and not the narrow jet episodes, in the wind used to compute the PPNe in the next section. 
The narrow  jets could have an observational counterpart, as we will discuss in \S~5.

\subsection{Model AS. The launching of  the wide X-wind }

The launching in Model AS (Figure 9) is similar to Model AF, but since the inner boundary lies at a larger radius (0.04 AU), the launching is less intense, 
because the escape velocity at the inner disk has
decreased from 517 to 206 \kms. The decrease of the escape velocity facilitates the release 
of magnetic pressure accumulated at the inner edge of the disk due to the centrifugal barrier.

The wind is launched between 20 and 50 degrees from the polar axis. Most of the wind momentum lies in
this region as well as the position-dependent mass-loss rate, as is shown in Figure 10. 
The wind also forms several pulses like model AF, but the time scale for the episodic events is longer
(weeks instead of days). Some parameters extracted from the final model are shown in Table 4.

A narrow jet is also formed, with a velocity of $\sim 250$ \kms, from the collision
of the gas with the polar axis (Figure 9, left panel). This jet, however, lasts longer than the one
that forms in model AF and is more stable in time. The mass loss rate is also very small
($\sim 10^{-12}$ \Moy).

\subsection{The formation of proto-planetary nebulae by wide jets}

We have seen above that the computed magnetic wind has two main components: the wide jet, which
carries most of the momentum, and the narrow jets, which are of short duration, without any
important contribution to the momentum. We now proceed to compute the effect that the
wide jet has on the circumstellar medium. 
Although the wide jet is composed of pulses, we focus on its time-averaged behavior since the time between pulses is very short (a few days) in comparison with the characteristic time scale of years simulated in the much larger computational domain.

Following the scheme of paper I, we adopt several scenarios for the circumstellar medium
previous to the CE event. Since the wind of the precursor AGB star can be in any of the phases of the thermal pulses at the moment of CE, the rates 
of mass loss span values corresponding to high-states with $\Mdot = 10^{-6}$ \Moy\ (Models B6 and C6), low-states with $\Mdot = 10^{-9}$  \Moy\  (Models B9 and C9), and intermediate  values of $10^{-7}$ and $10^{-8}$ (Models B7, C7, B8, and C8). We also compute one case where the precursor was a RG star (Model D9). In each of these runs we apply the time-averaged wide jet from Model AF as an inflow condition on the inner boundary.

As discussed in \S~3, the uncertainty
in the magnetic field evolution in the wide jet is taken into account by assuming that the toroidal magnetic field 
drops as  $1/r$ 
in Models B$x$ (option a, stronger hoop stress) at the
inner boundary, while Models C$x$ assume a drop by
 $1/r^2$ (option b, weaker hoop stress). 

Figure 11 shows the resulting nebulae for Models B6, B7, B8, and B9 at 50 and 1,000 years,
while Figure 12 shows the result for Models C6, C7, C8, and C9.

The effect of the AGB mass-loss rate on the nebula morphology is clearly seen in Figure 11. 
Higher AGB mass-loss rates (Model B6) imply a larger thermal pressure in the circumstellar medium 
and a larger ram pressure, which opposes 
the expansion of the outer shock, producing
narrow shapes. Lower AGB mass-loss rates (Model B9) produce wider
morphologies. The effect is also important in the collimation of the wide jet.
A larger AGB mass-loss rate translates into a larger compression of the shocked gas
in the interior of the nebula, which is dominated by magnetic pressure in these
cases. This also has important effects on the confinement of the jet, since
the gas is forced to 
lie close to the polar axis for the cases with larger
pressure (Model B6). In contrast, lower AGB mass-loss rates produce nebulae 
with lower inner pressure, which allows the expansion of the jet in the $\theta$ direction (Model B9).

The differences in the treatment of the toroidal magnetic field
are also 
notable in comparing Figures 11 and 12.  A larger contribution 
of the magnetic pressure and hoop stress also  produces 
greater collimation 
of the jets (Figure 11). 

We note that there are many similarities with the study by 
Balick et al.\ (2020). For example,
their solutions with lower jet densities are equivalent to our solutions with
higher AGB mass-loss rates, since it is the contrast between the pressure of the circumstellar medium and the interior of the nebula that is the important parameter
defining the resulting shape. The same trend applies to the intensity of the magnetic field.
Since our wide jet is the result of the computation of Model AF, we cannot 
vary the taper angle as in Balick et al.\ (2020). If we make a comparison with a
tapered flow, we think that our wide jet behaves similarly to those with 
an angle of $20^{\circ}$ 
as deduced by the variation of density and radial velocity with co-latitude (see Figure 6).

Figure 13 shows the case of Model D9 at a young stage, 276.7 days after the CE event. At this time, the  physical dimensions of the model give the
best spatial resolution. After this time, the expansion of the grid 
takes place 
and the spatial resolution decreases. 
At this young stage 
the ejected envelope in the equatorial regions is still visible. However, the 
ejected envelope is overtaken tens of years later. Note that Model D9 has the fastest expansion
velocity of the ejected envelope ($\sim 40$ \kms). In Models B$x$ and C$x$, with half the velocity, the ejected envelope is surpassed in a shorter time, 
leading to the lack of visibility of the ejected envelope
in Figures 11 and 12.

It is interesting to note that the shocked gas inside the nebula is sub-Alfv\'enic
(green-yellow-red colors in the right-top plot in Figure 13), i.e., the magnetic pressure 
is dominant in the dynamics, but there is a transition to the 
super-Alfv\'enic regime
in the polar region. Also, it is noteworthy
that the collimated jet near the polar axis is very cold 
(right-bottom plot in Figure 13) and approaches the 
cutoff  temperature imposed in our radiative
cooling scheme of 100 K. As a consequence, this
jet should be in a molecular phase. On the other hand, the shocked gas is above $10^4$ K in some places and could cool via the emission of optical lines.

For comparison, Table 5 shows the total kinetic energy acquired by the outflows in Models B$x$ and C$x$. 
The given values are after 1,000 years of evolution. Values of $4-5 \times 10^{45}$ erg   
are in good agreement with the observed values reported by Bujarrabal et al.\ (2001). Beside the 
nebular shape, this is probably the best observable with which to test our computed wide jet, since 
observational measurement of the mass-loss rate can usually only be done indirectly.
Note that lower values of
the kinetic energy correspond to 
a greater
slowdown by the circumstellar medium, proportional to the AGB mass-loss rate.

\subsection{The formation of proto-planetary nebulae by wide X-winds }

The X-wind produced in Model AS is introduced as an inflow condition
at the inner boundary for Models E$x$ (option a, larger hoop stress) and F$x$ (option b, smaller hoop stress).
The results are displayed in Figure 14. 

The results are quite different with respect to models in the B, C, and D series for two reasons.
First, the momentum of the wind lies in an angular extent ranging from 20 to 
50 degrees from the polar axis, while in Model AF the wind momentum 
is distributed in the cone from 0 to 20 degrees. This produces shapes that are
more elliptical in the case of Models E$x$ and quadrupolar
in the case of F$x$.  The elliptical shapes in Models E$x$ are produced because the
formed nebulae are subject to a larger magnetic pressure that redistributes the gas in all
directions, moving the gas either toward the pole or toward the equator. Models
F$x$, on the other hand, form quadrupolar shapes since in these models the winds behave more ballistically.

Note that in general,
the dynamics is extremely radiative, due to the high densities and low expansion velocity of
the X-wind ($\sim 30 $ \kms). 

The quadrupolar shapes in Models F$x$ are axisymmetric in 2D. However, in 3D their shapes might be different, since either the non-linear thin-shell instability (Vishniac 1994) or variations
in the launching in the $\phi$ direction can break this symmetry.

\section{Discussion}

Two basic mechanisms for the launching of jets have been discussed in the literature.
In the first mechanism, Blandford \& Payne (1982) proposed the launching by a magnetocentrifugal effect. Here, the magnetic
twist created by the rotation of gas in the disk drives out mass via the ${\bf J} \times {\bf B}$ forces on the
field lines that leave the disk surface and extend to large distances.
The second relies on the increase of the magnetic pressure from the toroidal magnetic field or plasma-gun launching (Contopoulos 1995). This is also called a Poynting flux-dominated or magnetic tower jet in the literature (Huarte-Espinosa et al.\ 2012; Sabin et al.\ 2020).

In each of the launching scenarios, angular momentum is extracted from the inner disk, allowing gas to accrete onto
the central object.  The torque exerted on the disk material could be external to the disk (Blandford \& Payne 1982) or internal to the disk (Balbus \& 
Hawley 1991).
Both types of torques were studied by
Stone \& Norman (1994), and our work is based on this last study. 
Certainly, there is a need to carry out much more work in order to explore this complex scenario for
the magnetic launching in PPNe. This will be the subject of future investigations.
Note that in Models AF and AS, we only accounted for the toroidal magnetic field
in the ejected envelope. Including the poloidal component in the ejected envelope 
(Ohlmann et al.\ 2016) could be important for producing a strong 
magnetocentrifugal effect as well.

We note that in this study, the wide jet originated in the circumbinary
disk, but this does not prevent the possible formation of a second jet 
originating from
a smaller accretion disk around the secondary star, since the computed
mass-accretion rate at the inner boundary is $\sim  4 \times 10^{-5} $ \Moy.

Several  observed PPNe  can be discussed in the light of these new calculations. We focus on the following three PPN.

{\bf W43A}.  This is a water fountain nebula observed recently with ALMA by Tafoya et al.\ (2020).  
It has an estimated age of only 60 yr
and is the best example ever observed, with an unprecedented resolution,
of a molecular jet in a PPN  emanating from the central source.
The jet has an expansion velocity of 150 \kms, and it shows a small deceleration 
at both ends. 
The jet is formed by a sequence of blobs, with 
each blob separated by a time interval of 5-7 yrs.
 
A possible explanation for the origin of the blobs (Tafoya et al.\ 2020) is that they
are formed in successive passages by a secondary star in an eccentric orbit.
Thus, the interval between blobs would be equal to the orbital period of the binary system.
This scenario requires that the binary system has not undergone 
a CE event.

However, in view of our results, we suggest an alternative
explanation.
Figure 15 shows W43A  in comparison with Model B8  at 50 yr of evolution (left panels). 
Model B8 is also shown  in Figure 16 with different physical quantities in detail.  Although the comparison is suggestive, it should be pointed out that the jet in Model B8 is smooth, without blobs 
due to the fact that we have not implemented any variability in the injected wide jet (although we found
that the solution for the wide jet in Model AF consisted of pulses). 
Furthermore, the 2-D axisymmetric computations do not allow one to compute properly the different instabilities originating in a magnetically confined jet (e.g., kink instabilities)
which could result in the formation of blobs as well as a deceleration of the jet. 

However, we note that there are astronomical 
observations of other objects (young stellar objects) which show that  jets 
have knot or blob-like structure, with
a similar time interval between blobs (2-6 years in the case of the T-Tauri star RW Aur A 
studied by Takami et al.\ 2020).  Thus, it is possible that W43A has already 
evolved beyond the CE event
60 years ago, which triggered the jet and the formation of the PPN. If this is the case, the
jet is formed by the circumbinary disk and not by the disk around the 
companion. 

A second clue in favor of the post-CE scenario is the fact that the nebula shows
point-symmetry in the lobes. Usually, since point-symmetry was discovered (Corradi \& Schwarz 1995), 
it has been found to be associated with a  precessing jet. However, in this case it is clear that the central jet is not precessing. 
Thus it is difficult to explain the point-symmetry using a single jet, which would be the case
of a binary in an eccentric orbit. However, as we mentioned earlier, 
there is a large
amount of gas ($\sim  4 \times 10^{-5} $ \Moy) that 
flows toward the binary and 
forms a secondary accretion disk around the companion with the possibility of launching a 
secondary jet. This jet could be wobbling (Livio \& Pringle 1997) and could be responsible for the point-symmetry observed in W43A.

A third clue is that there are no previous signs of passages by the binary earlier than
60 years, and
it is not clear why the first passage that formed a jet started 60 years ago. In the case of the post-CE
scenario, it is clear that the event that triggered the formation of the nebula and the jet was the ejection of the envelope.

{\bf OH231.8+4.2}.  The Calabash nebula is an extremely interesting object, also recently 
observed with ALMA (S\'anchez-Contreras et al.\ 2018). We do not  attempt to 
provide a 
detailed explanation  for the origin of the nebula (see for example Balick et al.\ 2017; Sabin et al.\ 2020), but we
will focus on one interesting feature, which is the emission in H$\alpha$ studied by
Bujarrabal et al.\ (2002) with HST. Figure 15 shows our Model C9 at 1,000 yr in 
comparison with OH231.8+4.2 (right panels), which S\'anchez-Contreras et al.\ (2018) estimate to have an
age of 800-900 yr. Model C9 is also shown in detail in Figure 17. 
Following the notation by Bujarrabal et al.\ (2002), we found that the features 
C1, C2, C3, C4, E1, and E2
can be associated with the reverse shock, since the density and temperature are large 
according with Model C9 (Figure 17), while feature D is associated with the forward shock. 
This confirms the original explanation by Bujarrabal et al.\ (2002). This is 
clearer 
in the region close to the fish bowls (S\'anchez-Contreras et al.\ 2018). Note that
the H$\alpha$ emission close to the center of the nebula originates at the inner 
edge of the fish bowls, in concordance with the reverse shock in Model C9, while the
outer shock is external to these features. The fish bowls 
form in Model C9
due to the interaction of the dense and slow wind ($\sim 30$ \kms) 
characterized by
$50^{\circ}-60^{\circ}$ of co-latitude angle (Figure 6) 
with the ejected envelope at the equatorial regions.

{\bf He 3-1475}.  The wind observed in this PPN has two main components: a fast wind flowing with velocities in the range 
$\sim 150 - 1200$ \kms  (Riera et al. 1995; S\'anchez Contreras \& Sahai 2001), and an ultrafast wind  moving much faster (up to $2300$ \kms) (S\'anchez Contreras \& Sahai 2001).    
The latter is highly collimated (length/width $\sim$ 7) and close to the central star, 
with a radially increasing velocity.
S\'anchez Contreras \& Sahai identify the ultrafast wind with a ``pristine'' post-AGB outflow that has not been altered by its interaction with the AGB envelope. 
These two types of winds are in line with our result in Model AF. The fast wind could be
identified  with the wide jet, while the ultrafast wind could be associated with the narrow jets
(we found velocities up to 3,000 \kms).
This is still rather speculative, and much more work is needed to understand this
dual 
behavior of the wind. Note that the velocities of the narrow
jets are much faster 
than the escape velocity of the binary system. We also cannot  discard the idea that the 
ultrafast wind could be associated with a second jet formed by the companion star.

Nebulae like M 1-63 could be formed from models like E8 or E9, while other elliptical
nebulae which present the characteristic ansae and flyers like NGC 6826 could be formed 
from models like E6, E7, or E8.  On the other hand, nebulae like He 2-47, He 2-339 
and IRAS 19024+0044 could have originated from models like F6 and F7.

The last point to discuss is the lifetime  of the circumbinary disk.
According to Model AF, the mass in the disk is $M_{\rm disk}=0.515$ \Mo \ (Table 3). 
Combining 
the mass accreted at the inner boundary of $\Mdot_{\rm acc} = 4 \times 10^{-5}$ \Moy\ and the mass-loss rate of the wide jet of $\Mdot = 1.3 \times 10^{-7} $ \Moy\ gives a total
disk survival time of 12,800 yr, assuming that the accretion and ejection rates are constant in time. 
There is observational evidence for the presence of circumbinary disks around
post-AGB stars. For example, Bujarrabal et al.\ (2018) using ALMA shows a disk around the post-AGB
IRAS 08544-4431, with a derived lifetime of $\sim 10,000$  yr.

This survival time is of the same order of magnitude as the PN lifetime. 
This is interesting in the context of the total number of PPNe$^*$ and PNe$^*$, where
PPNe$^*$ would correspond to 
those PPNe that have winds that cannot be explained by 
radiation forces and are formed by jets, and PNe$^*$ would correspond to those PNe that resulted from a CE event.
The number ratio of PPNe$^*$/PNe$^*$ could give information regarding the efficiency of
the ejection at the CE event.
In our case, the total ejected envelope mass was $ 0.172$ \Mo\ (25\%), and the
total envelope  mass  that remained bound was $ 0.517$ \Mo\ (75\%).  
In the extreme case that all the envelope mass was ejected, there would be no mass available
 to form the circumbinary disk, so the transition to PN$^*$ would be direct as in paper I,
without passing through a PPN$^*$ phase. So the ratio PPNe$^*$/PNe$^*$ would be 0 in this case. 

The above survival time is likely to be an overestimate since 
 other processes can facilitate the dispersal of the gravitationally bound matter in the 
 system in the form of a circumbinary disk. In particular, the line driven wind from the 
 remnant stellar core as well as the photoevaporation of the disk produced by the stellar
 radiation field for a sufficiently hot core can be effective in its dissipation.  
 This naturally will terminate the accretion phase and the launching of jets, which are 
 important topics for future studies.

A final remark is that we have not considered the formation of dust in the
stages after the CE (Glanz \& Perets 2018; Iaconi et al. 2020). Unfortunately, 
we cannot work properly with dust in ZEUS-3D, since it does not include
a multi-fluid scheme like other codes. Dust is important not only 
in driving cool winds by radiation pressure, but also in the detection of magnetic
fields by dust grain alignment. 
This probably is the most promising probe to study the magnetic field and to confront models
since ALMA allows to observe linear polarization in PPNe in great detail.
This is a challenging work that could be the subject of a future study.

\section{Conclusions}

We have computed successfully the launching of two morphologically different magnetic winds from a circumbinary 
disk formed after a common envelope event. The winds are  non
steady, and they are launched in the form of small bursts, 
with periods of days or weeks.  The winds can be described as a wide jet with a maximum radial velocity of $\sim 230$ \kms\ and an average  mass-loss rate of  $\sim 1.3 \times 10^{-7}$ \Moy\ in the case of remnant systems with orbital periods $\lesssim 2$  days or a X-wind with a velocity of $\sim 30$ \kms\ and a mass-loss rate of $\sim 1.67 \times 10^{-7}$ \Moy\ for systems with orbital periods $\sim 5$ days.
Although the short time scale variability of the wind has yet to be observed, 
future continuum monitoring of jets in PPNe with high spatial (1 AU) and temporal  (1 days/weeks) resolution will be required, which will be observationally challenging.

We have also shown the
type of nebulae that could result from the injection of the computed wide jet into the circumstellar medium. It has been demonstrated that 
the formed PPNe are characterized by bipolar, elongated shapes, which are in good agreement with observed PPNe, whose kinetic energies reach  $\sim 4 \times 10^{45}$ erg at 1,000 years.
On the other hand, the nebulae resulting from the injection of the X-wind are more elliptical and
quadrupolar in shapes, with smaller expansion velocities. 

In the future, we plan to investigate the launching of magnetic winds in 
greater detail, taking into account three-dimensional magnetohydrodynamical effects as well as the effect of the binary nature of the stellar components on the structure of the circumbinary disk.  Specific attention will be focused on the inner 0.2 AU which, is crucial in providing an understanding of the launching process.

\acknowledgments

We thank the referee for his/her comments which improved the manuscript.
We thank Michael L.\ Norman and the Laboratory for Computational
Astrophysics for the use of ZEUS-3D. The computations
were performed at the Instituto de Astronom\'{\i}a-UNAM at  Ensenada.
G.G.-S.\ is partially supported by CONACyT grant 178253.
Partial support for this work has been provided by NSF through grants
AST-0200876 and AST-0703950.

{\bf \software{ZEUS-3D (version 3.4; Stone \& Norman 1992, Clarke 1996)}}



\clearpage

\begin{table}
\begin{center}
\caption{Numerical models}
\begin{tabular}{llllll}
\tableline\tableline
Model &  Resolution & Description  & Log $\Mdot_{\rm AGB}$ ($\Moy$)& Option  &   Figures \\
\tableline
AF           & $200 \times 200$ & Fast Wind Launching & does not apply & &  1,2,3,4,5,6,7,8  \\
B6,B7,B8,B9  & $800 \times 200$ & Expanding PPN  & -6,-7,-8,-9   &  a & 11            \\
C6,C7,C8,C9  & $800 \times 200$ & Expanding PPN  & -6,-7,-8,-9   &  b & 12            \\
D9           & $800 \times 200$ & Expanding PPN  &          -9   &  a & 13            \\
\tableline
AS           & $200 \times 200$ & Slow Wind Launching & does not apply & &  9,10  \\
E6,E7,E8,E9  & $800 \times 200$ & Expanding PPN  & -6,-7,-8,-9   &  a & 14            \\
F6,F7,F8,F9  & $800 \times 200$ & Expanding PPN  & -6,-7,-8,-9   &  b & 14            \\
\tableline
\end{tabular}
\end{center}
\end{table}

\begin{table}
\begin{center}
\caption{Model AF. Initial conditions}
\begin{tabular}{lr}
\tableline\tableline
Variable &  Value (cgs)  \\
\tableline
Total mass in the grid & (0.516  \Mo) $1.026 \times 10^{33} $ \\
Central point mass   & (0.96 \Mo)  $1.909 \times 10^{33}$ \\
Kinetic energy in $\phi$-direction         & $3.635  \times 10^{46} $   \\
Kinetic energy in $r$-direction            & $2.285  \times 10^{43} $   \\
Total kinetic energy                       & $3.638  \times 10^{46} $   \\
Gravitational potential energy             & $-6.861 \times 10^{45} $   \\
Magnetic energy                            & $1.051  \times 10^{44 }$   \\
Maximum density                            & $1.181  \times 10^{-3} $  \\
Maximum velocity in $r$-direction          & $4.30.  \times 10^{6}  $ \\
Maximum velocity in $\phi$-direction       & $2.89   \times 10^{7}  $  \\
Maximum magnetic field in $\phi$-direction & $1.281  \times 10^{4}  $  \\
Maximum magnetic field in $\theta$-direction & $ 64.05   $  \\
Maximum magnetic field in $r$-direction     & $ 6.76 $ \\
\tableline
\end{tabular}
\end{center}
\end{table}

\begin{table}
\begin{center}
\caption{End of Model AF }
\begin{tabular}{lr}
\tableline\tableline
Variable &  Value (cgs)  \\
\tableline
Total mass in the grid & (0.515 \Mo) $ 1.024 \times 10^{33} $ \\
Central point mass   & (0.96 \Mo)  $1.909 \times 10^{33}$ \\
Kinetic energy in $\phi$-direction         & $3.710  \times 10^{46} $   \\
Kinetic energy in $r$-direction            & $ 4.346 \times 10^{43} $   \\
Total kinetic energy                       & $3.715  \times 10^{46} $   \\
Gravitational potential energy             & $-6.851 \times 10^{45} $   \\
Magnetic energy                            & $1.340  \times 10^{44 }$   \\
Maximum density                            & $1.743  \times 10^{-3} $  \\
Maximum velocity in $r$-direction          & $2.29   \times 10^{7}    $  \\
Maximum velocity in $\phi$-direction       & $3.37   \times 10^{7}  $  \\
Maximum magnetic field in $\phi$-direction  & $2.275 \times 10^{4}  $ \\
Maximum magnetic field in $\theta$-direction  & $91.3   $ \\
Maximum magnetic field in $r$-direction  & $ 1.058 \times 10^{3}  $ \\
Mass-accretion rate (inner boundary)  &  ($4.36 \times 10^{-5} $ \Moy)  $ 2.75 \times 10^{21} $ \\
Mass-loss rate  (outer boundary)      &  ($1.23 \times 10^{-7} $ \Moy)  $ 7.75 \times 10^{18} $  \\

\tableline
\end{tabular}
\end{center}
\end{table}

\begin{table}
\begin{center}
\caption{End of Model AS }
\begin{tabular}{lr}
\tableline\tableline
Variable &  Value (cgs)  \\
\tableline
Total mass in the grid & (0.513 \Mo) $ 1.020 \times 10^{33} $ \\
Central point mass   & (0.96 \Mo)  $1.909 \times 10^{33}$ \\
Kinetic energy in $\phi$-direction         & $ 3.398  \times 10^{46} $   \\
Kinetic energy in $r$-direction            & $ 3.185 \times 10^{43} $   \\
Total kinetic energy                       & $ 3.400  \times 10^{46} $   \\
Gravitational potential energy             & $-6.827 \times 10^{45} $   \\
Magnetic energy                            & $1.208  \times 10^{44 }$   \\
Maximum density                            & $1.27  \times 10^{-3} $  \\
Maximum velocity in $r$-direction          & $1.46   \times 10^{7}    $  \\
Maximum velocity in $\phi$-direction       & $2.63   \times 10^{7}  $  \\
Maximum magnetic field in $\phi$-direction  & $1.01 \times 10^{4}  $ \\
Maximum magnetic field in $\theta$-direction  & $79.9   $ \\
Maximum magnetic field in $r$-direction  & $ 1.35 \times 10^{2}  $ \\
Mass-accretion rate (inner boundary)  &  ($2.02 \times 10^{-5} $ \Moy)  $ 1.27 \times 10^{21} $ \\
Mass-loss rate  (outer boundary)      &  ($1.67 \times 10^{-7} $ \Moy)  $ 1.05 \times 10^{19} $  \\

\tableline
\end{tabular}
\end{center}
\end{table}

\begin{table}
\begin{center}
\caption{Kinetic energy gained by models}
\begin{tabular}{lc}
\tableline\tableline
Model &  Total kinetic energy (erg) at 1,000 yr \\
\tableline

B6 &    $ 3.64 \times 10^{45} $         \\
B7 &    $ 4.40 \times 10^{45} $         \\
B8 &    $ 5.04 \times 10^{45} $         \\
B9 &    $ 5.60 \times 10^{45} $         \\
\tableline
C6 &    $ 2.08 \times 10^{45} $         \\
C7 &    $ 3.28 \times 10^{45} $         \\
C8 &    $ 4.18 \times 10^{45} $         \\
C9 &    $ 4.74 \times 10^{45} $         \\
\tableline
\end{tabular}
\end{center}
\end{table}

\clearpage

\begin{figure}
\epsscale{1.20}
\plotone{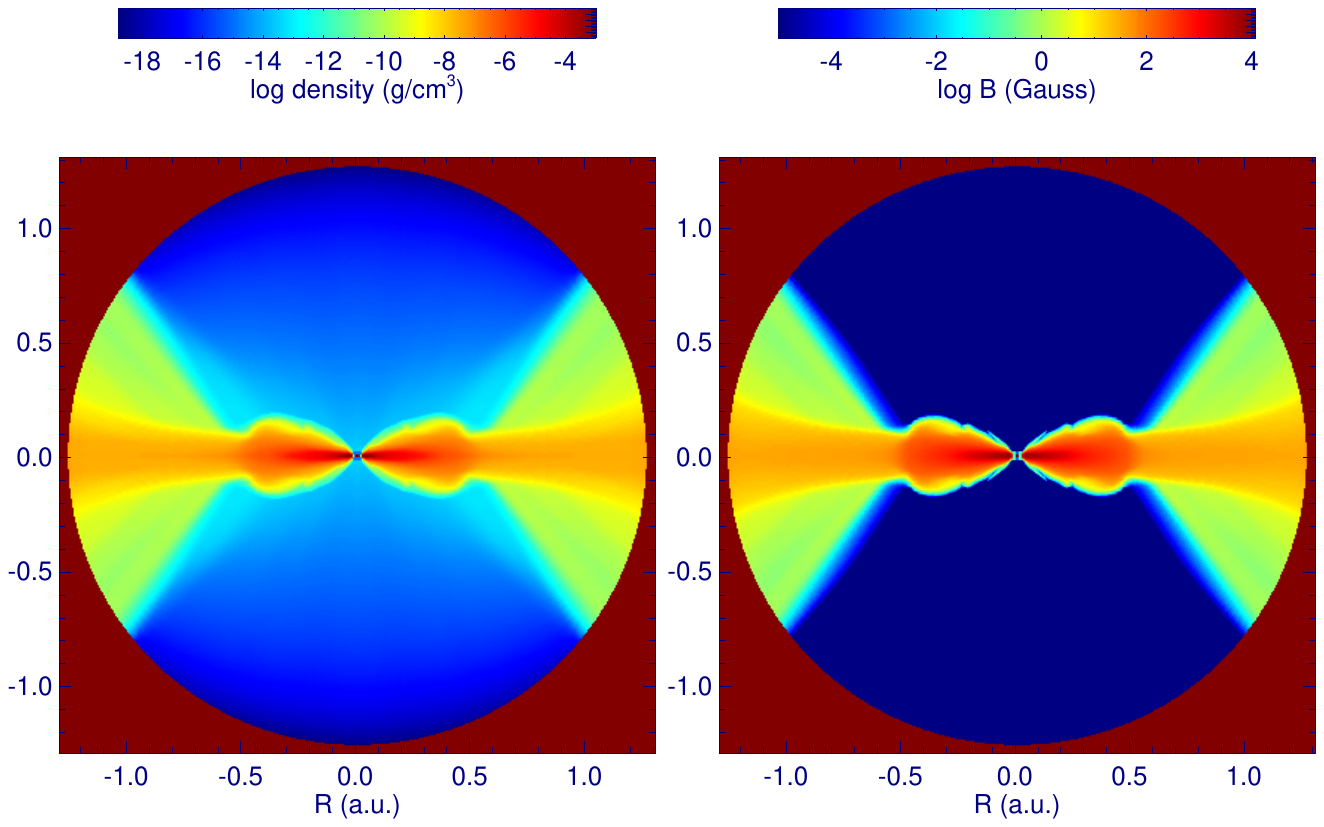}
\caption{Initial conditions for Model AF. Gas density and  magnetic field of the formed circumbinary disk at 156.7 days after the CE event.
The equatorial plane is perpendicular to the figure and  horizontally oriented. The symmetry 
axis is an imaginary, vertical line that passes through the center. The total mass of the gas in the grid is 0.516 \Mo . The magnetic field is toroidal, perpendicularly oriented to the figure.}
\label{f1}
\end{figure}

\begin{figure}
\epsscale{1.25}
\plotone{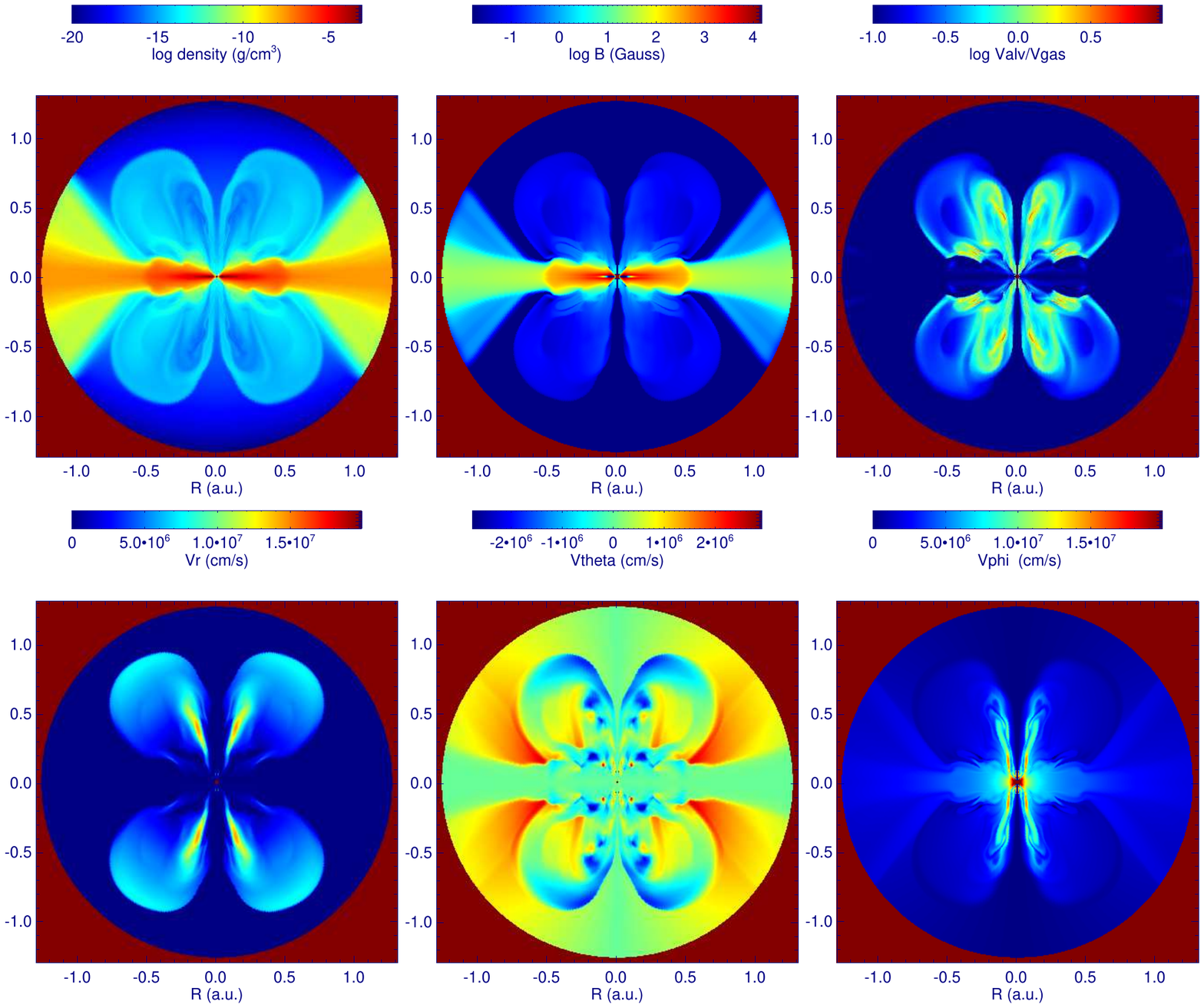}
\caption{Initial launching of the magnetic wind of Model AF. 
Gas density, toroidal magnetic field, ratio of the Alfv\'en to the gas velocity, radial velocity, co-latitude velocity, and rotational velocity, at 20 days from the initial conditions (176.7 days after the  CE event).}
\label{f2}
\end{figure}

\begin{figure}
\epsscale{1.25}
\plotone{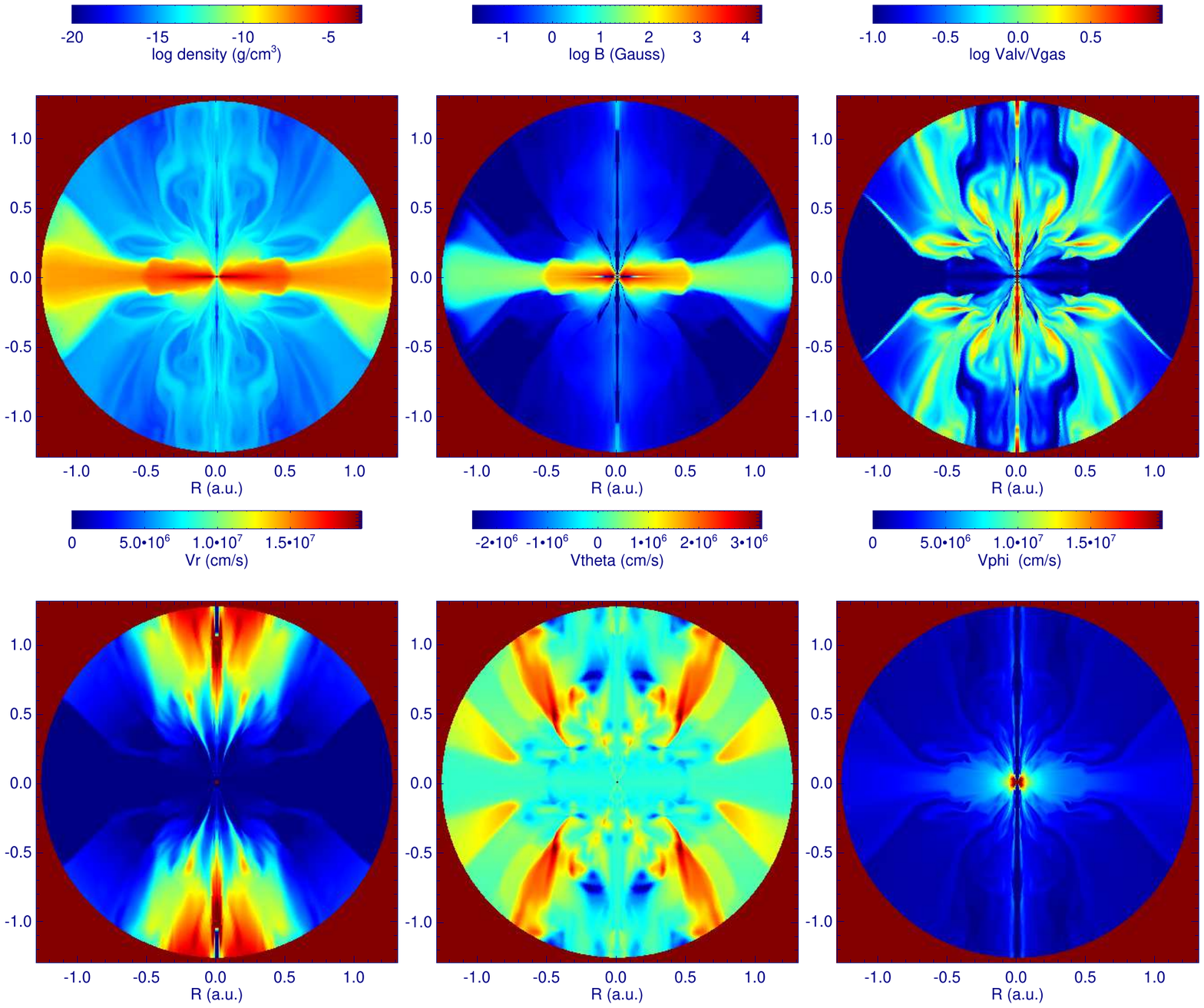}
\caption{Full developed magnetic wind of Model AF.
Gas density, toroidal magnetic field, ratio of the Alfv\'en to the gas velocity, radial velocity, co-latitude velocity, and rotational velocity, at 44 days from the initial conditions (200.7 days after the  CE event).}
\label{f3}
\end{figure}

\begin{figure}
\epsscale{1.25}
\plotone{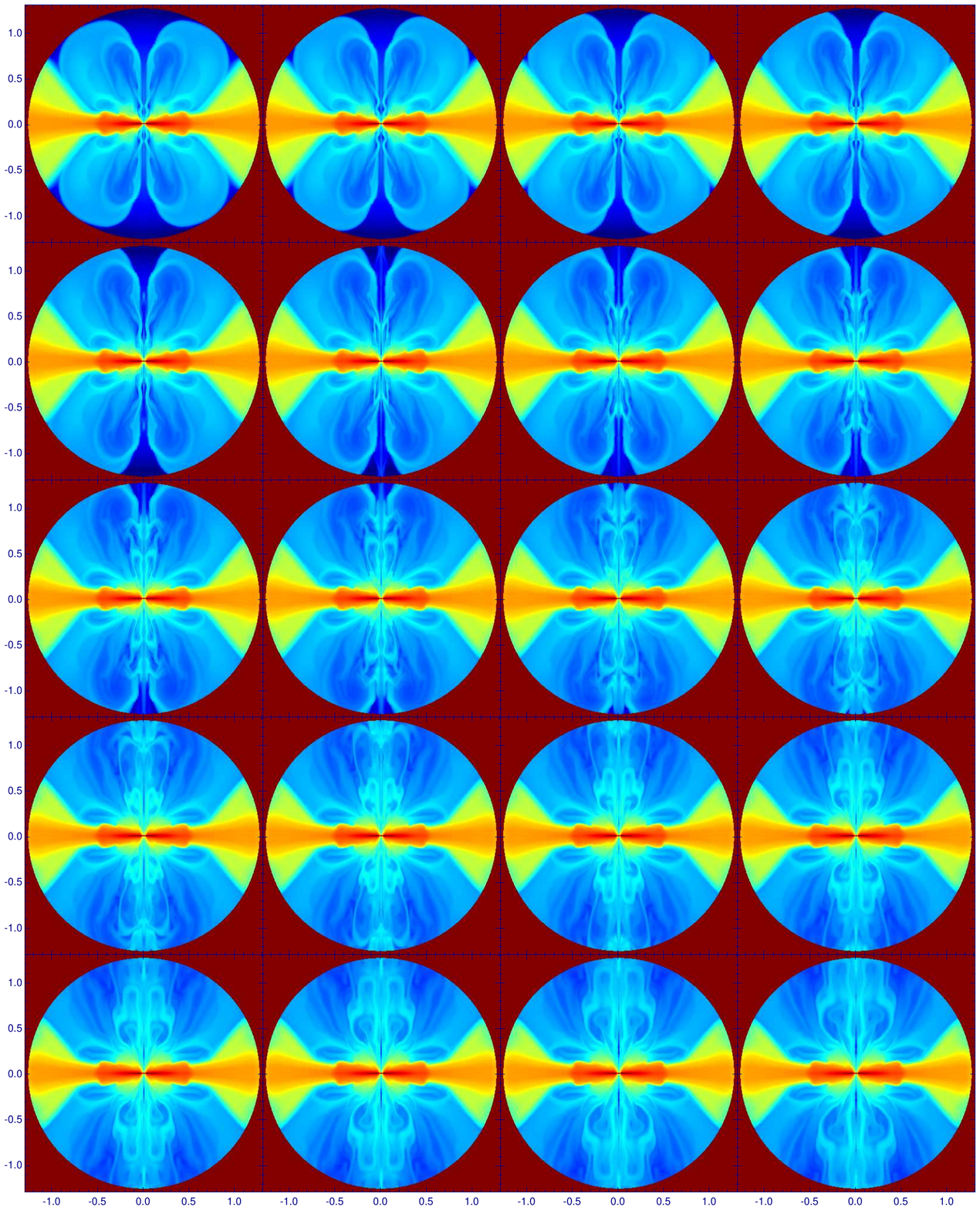}
\caption{Wind history of Model AF. Gas density snapshots with a time  interval of 1 day, starting at 181.7 days after the CE event. The last snapshot (200.7 days)
corresponds to Figure 3, with the same color scale. 
}
\label{f6}
\end{figure}

\begin{figure}
\epsscale{1.30}
\plotone{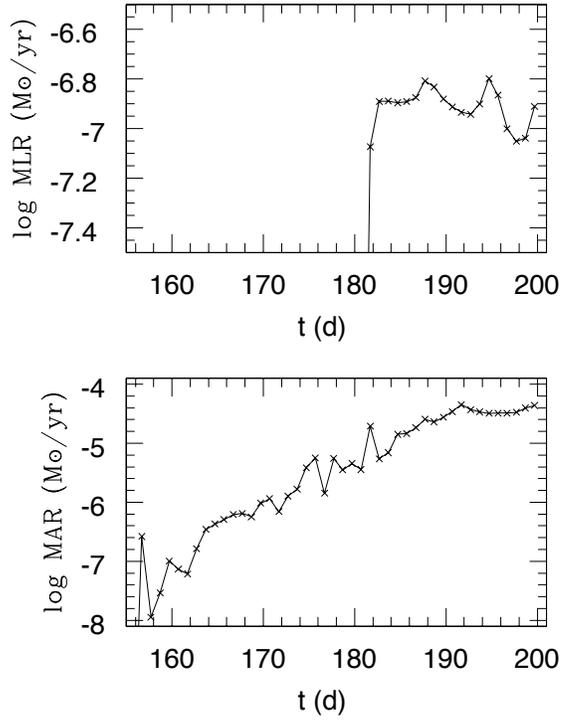}
\caption{History of the total mass-loss rate (top) and mass-accretion rate (bottom) of Model AF. The last point corresponds to 44 days from the initial conditions, 200.7 days after the  CE event.}
\label{f5}
\end{figure}

\begin{figure}
\epsscale{1.30}
\plotone{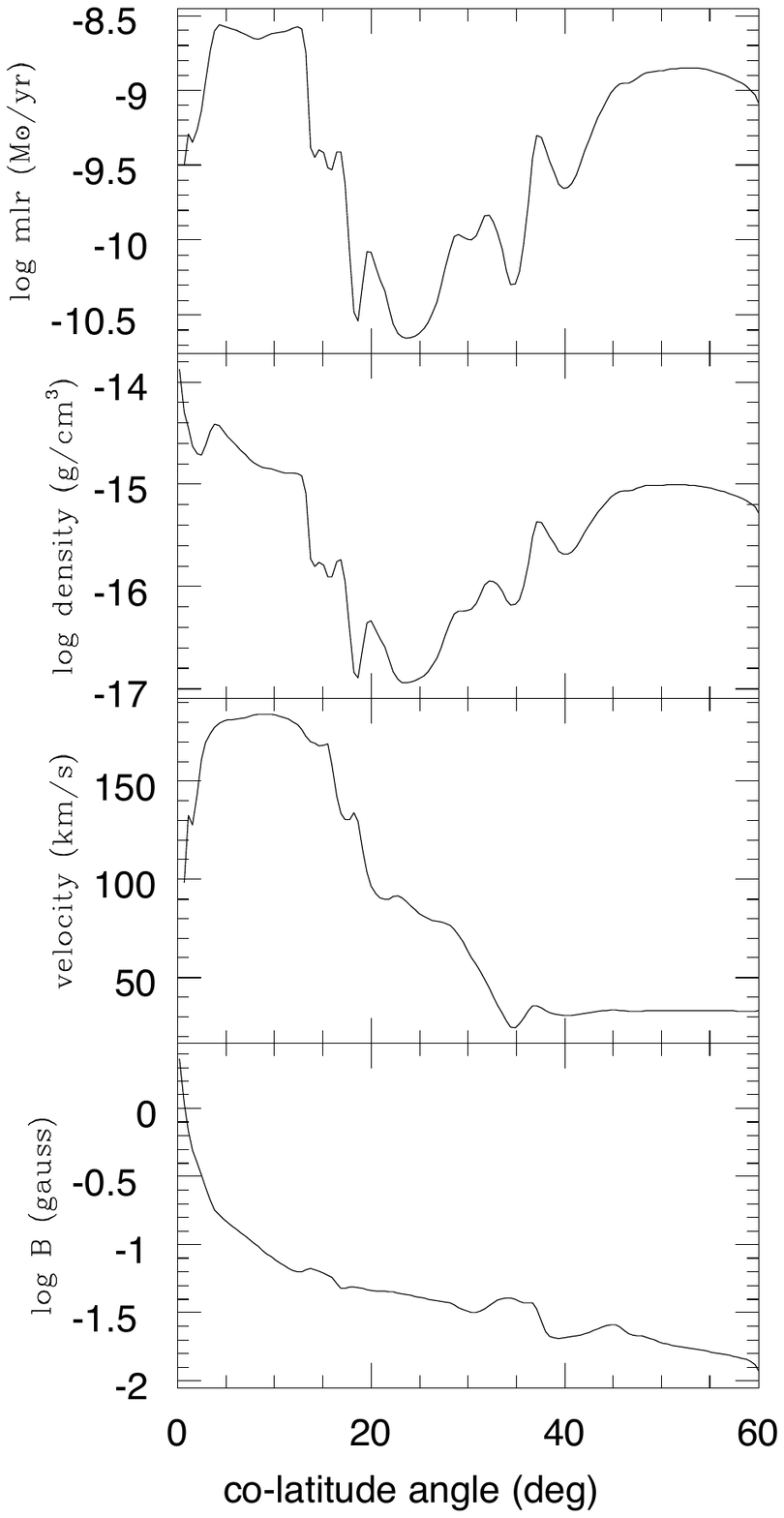}
\caption{Position-dependent mass-loss rate, gas density, radial velocity, and toroidal magnetic field at the outer boundary of Model AF, at 44 days from the initial conditions (200.7 days after the  CE event).}
\label{f4}
\end{figure}

\begin{figure}
\epsscale{1.25}
\plotone{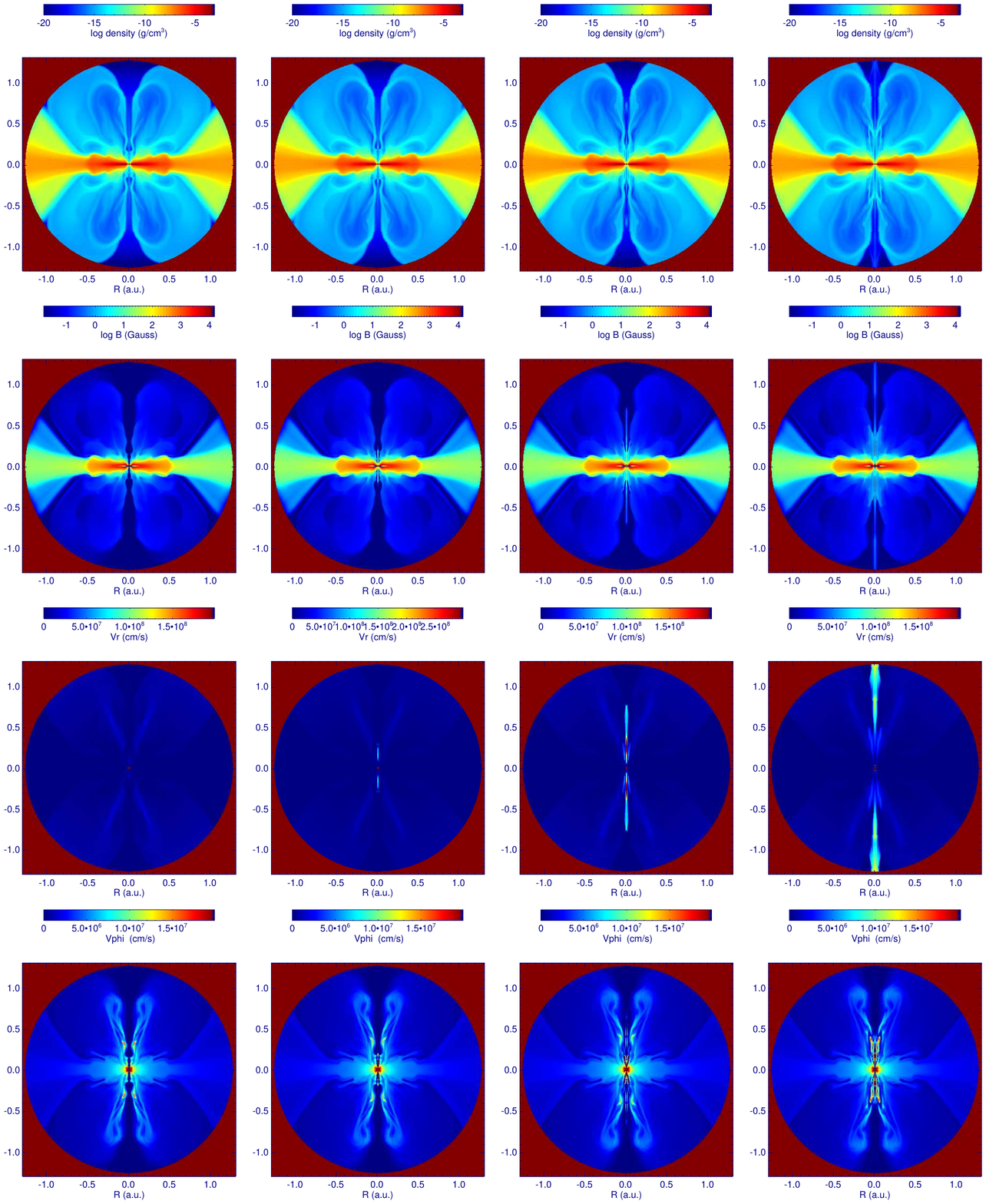}
\caption{Launching of a narrow jet. Gas density, toroidal magnetic field, radial velocity,
and rotational velocity snapshots of Model AF at days 
27, 28, 29 and 30 (183.7, 184.7, 185.7 and 186.7 days after the CE event).}
\label{f7}
\end{figure}

\begin{figure}
\epsscale{1.25}
\plotone{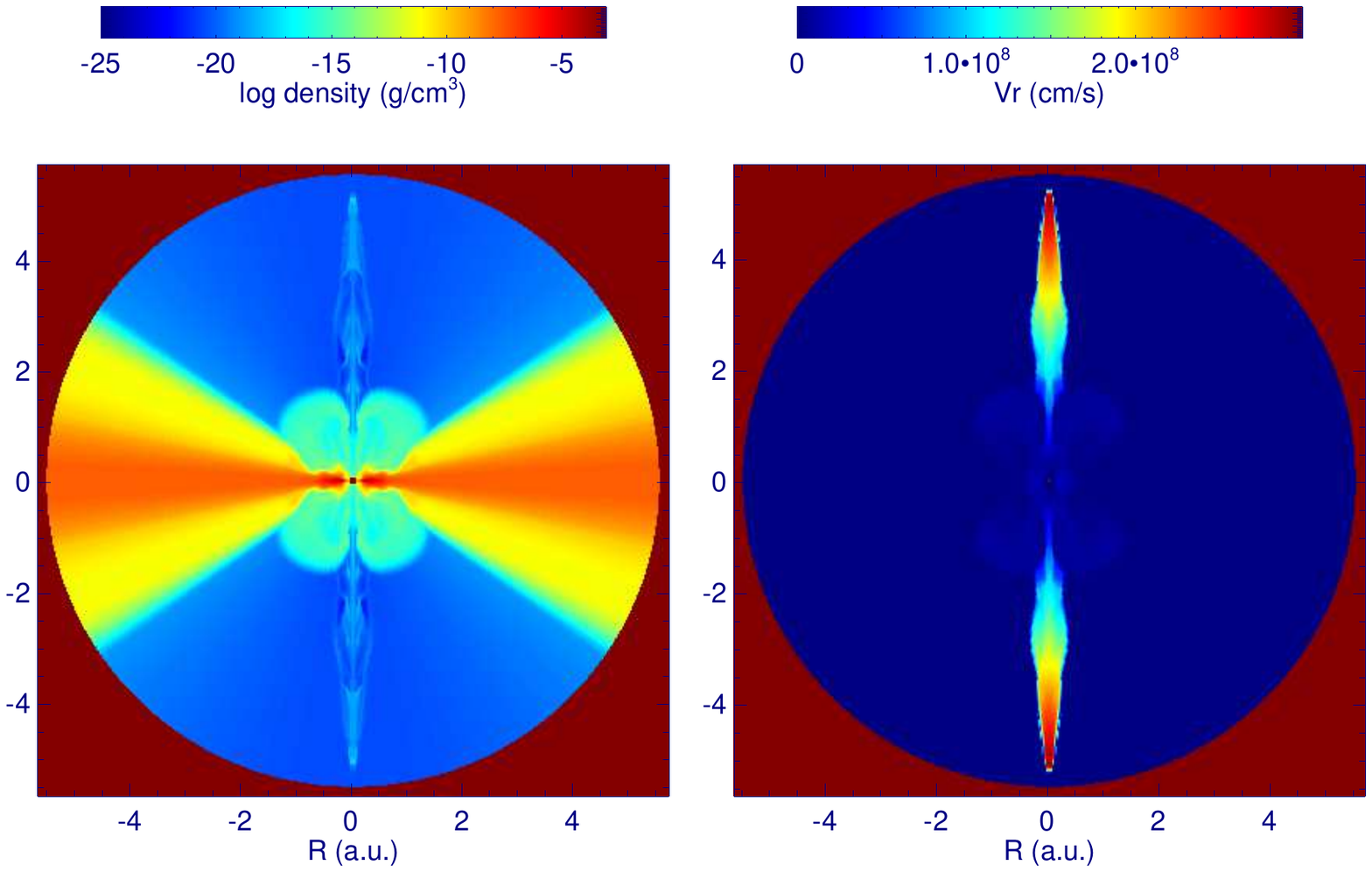}
\caption{Gas density and radial velocity snapshots of the follow up calculation of the narrow jet in Model AF at 188.7 days after the CE event. The grid has been expanded to track the bow shock of the jet.}
\label{f8}
\end{figure}

\begin{figure}
\epsscale{1.25}
\plotone{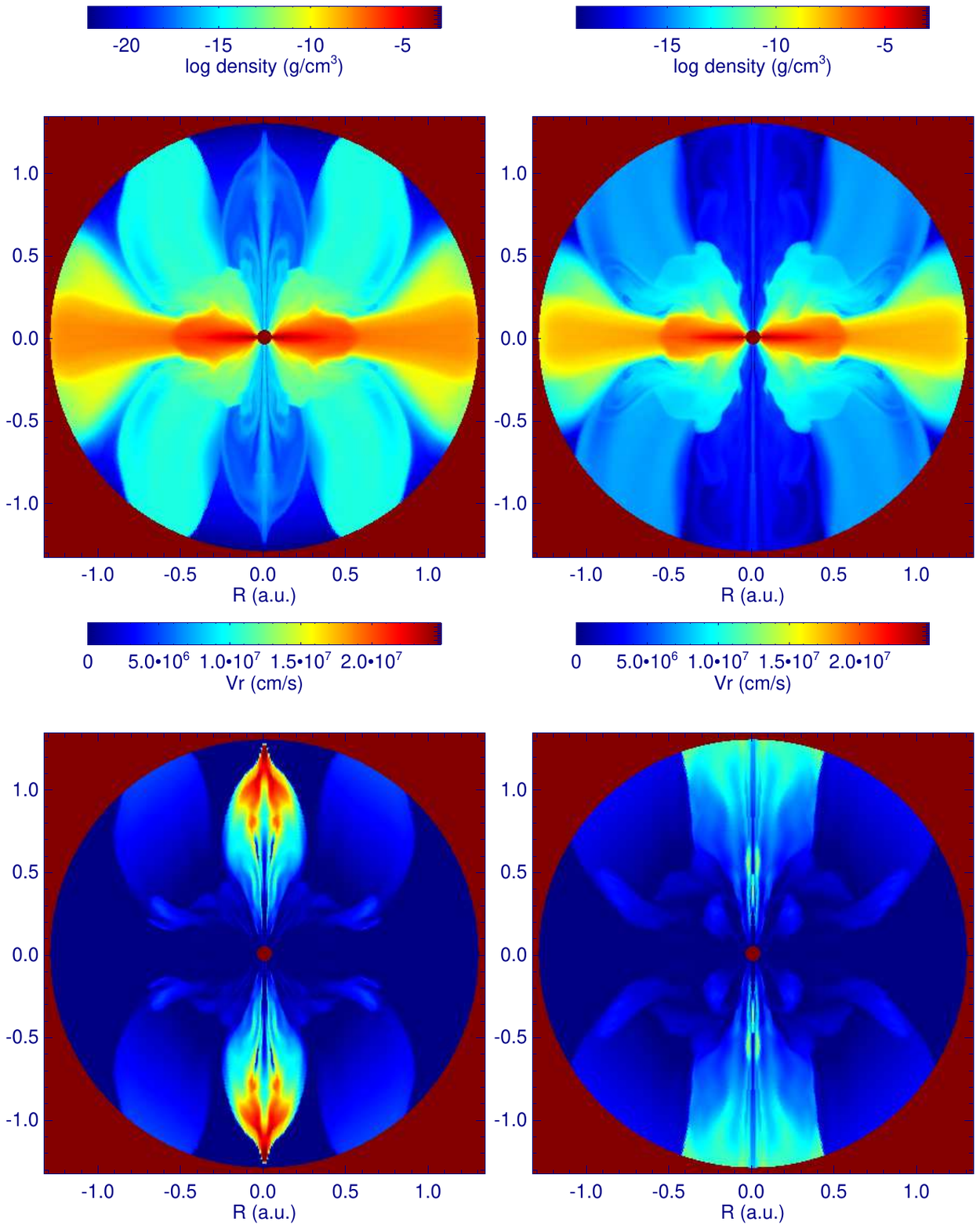}
\caption{Gas density and radial velocity of Model AS, at 45.4 days (left) and 56.9 days (right) from the initial conditions (202.1 and 213.6 days after the  CE event).}
\label{f9}
\end{figure}

\begin{figure}
\epsscale{1.25}
\plotone{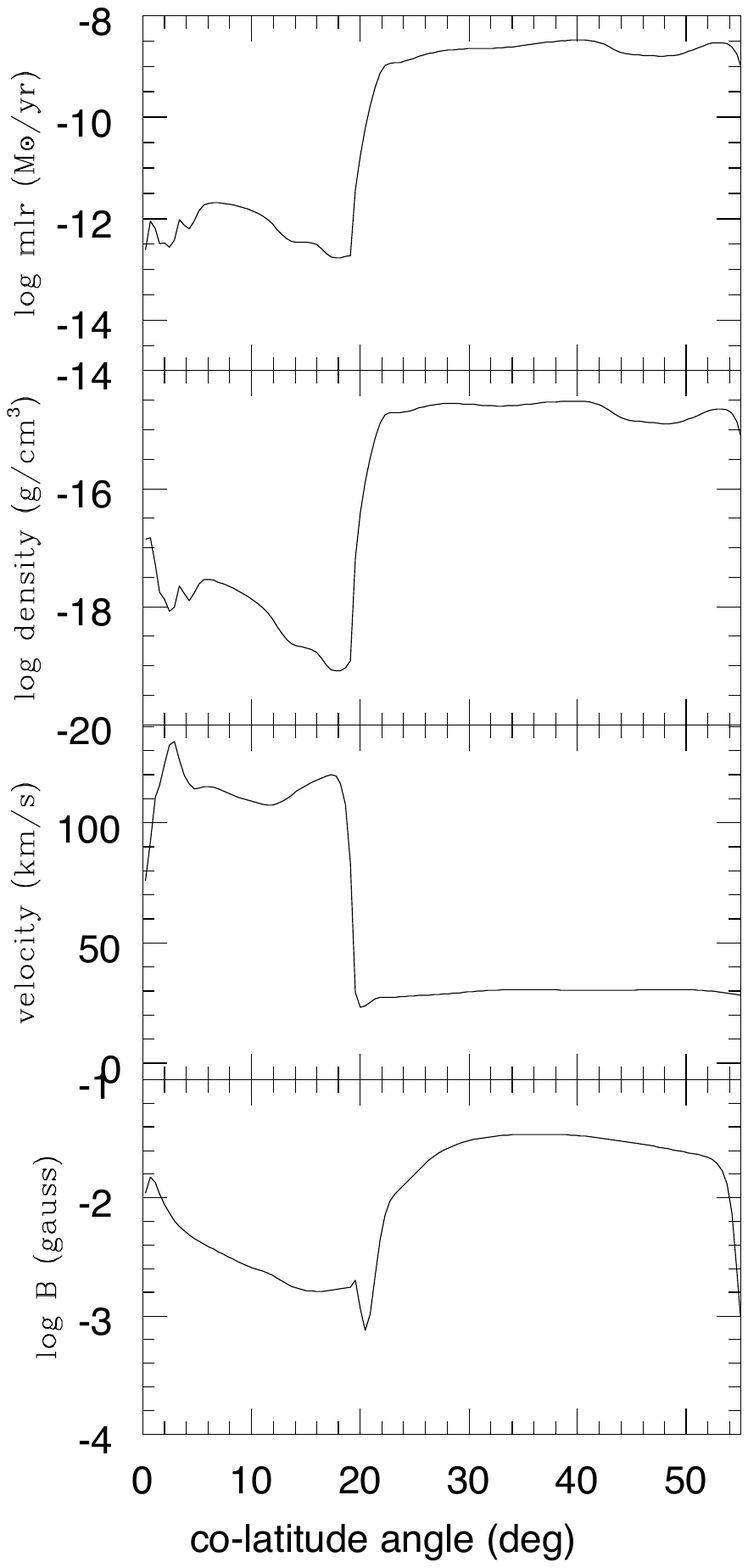}
\caption{Position-dependent mass-loss rate, gas density, radial velocity, and toroidal magnetic field at the outer boundary of Model AS, at 56.9 days from the initial conditions (213.6 days after the  CE event).}
\label{f10}
\end{figure}

\begin{figure}
\epsscale{1.25}
\plotone{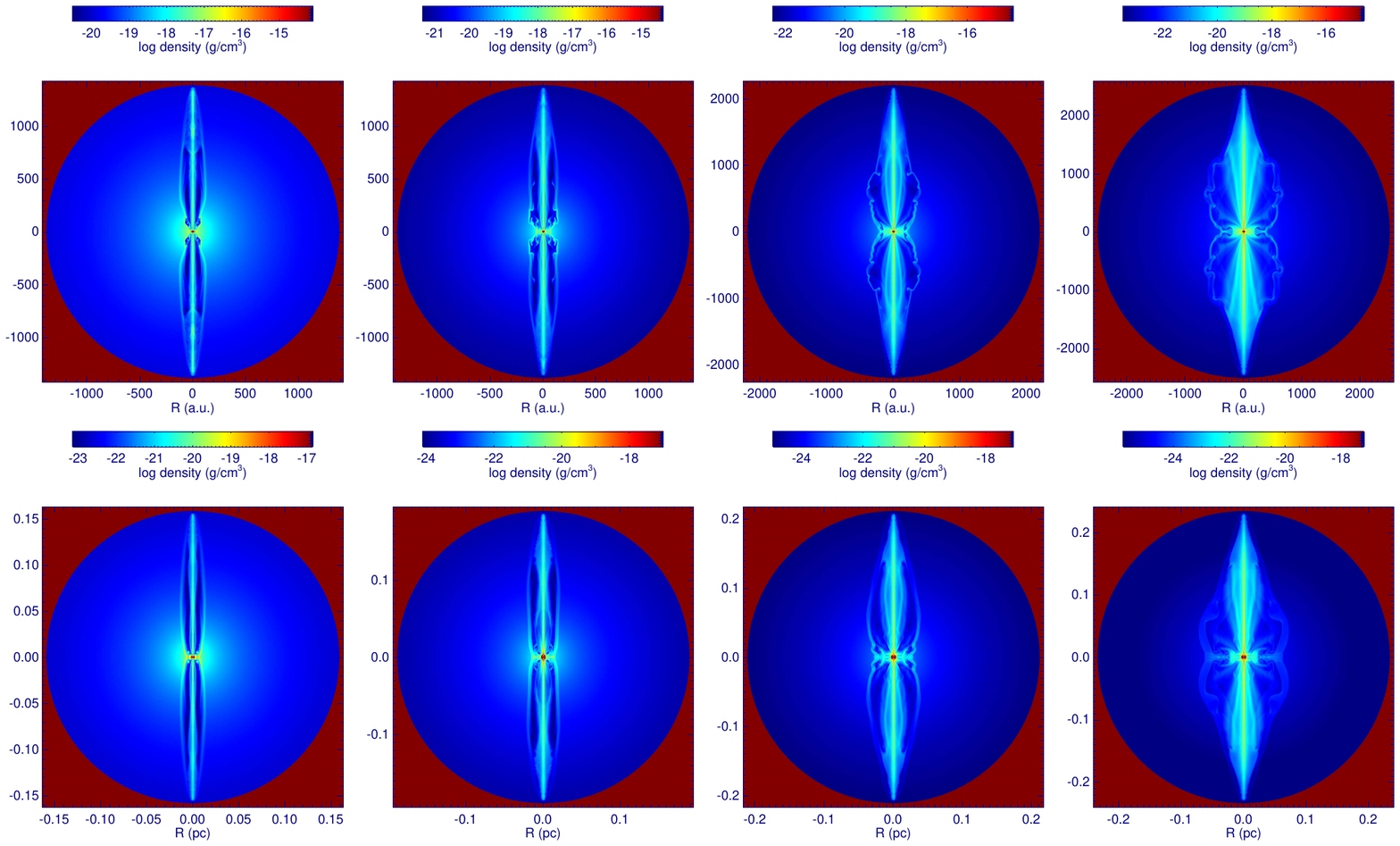}
\caption{Gas density snapshots at 50 yr (top) and 1,000 yr (bottom) of Models B6, B7, B8, and B9. }
\label{f11}
\end{figure}

\begin{figure}
\epsscale{1.25}
\plotone{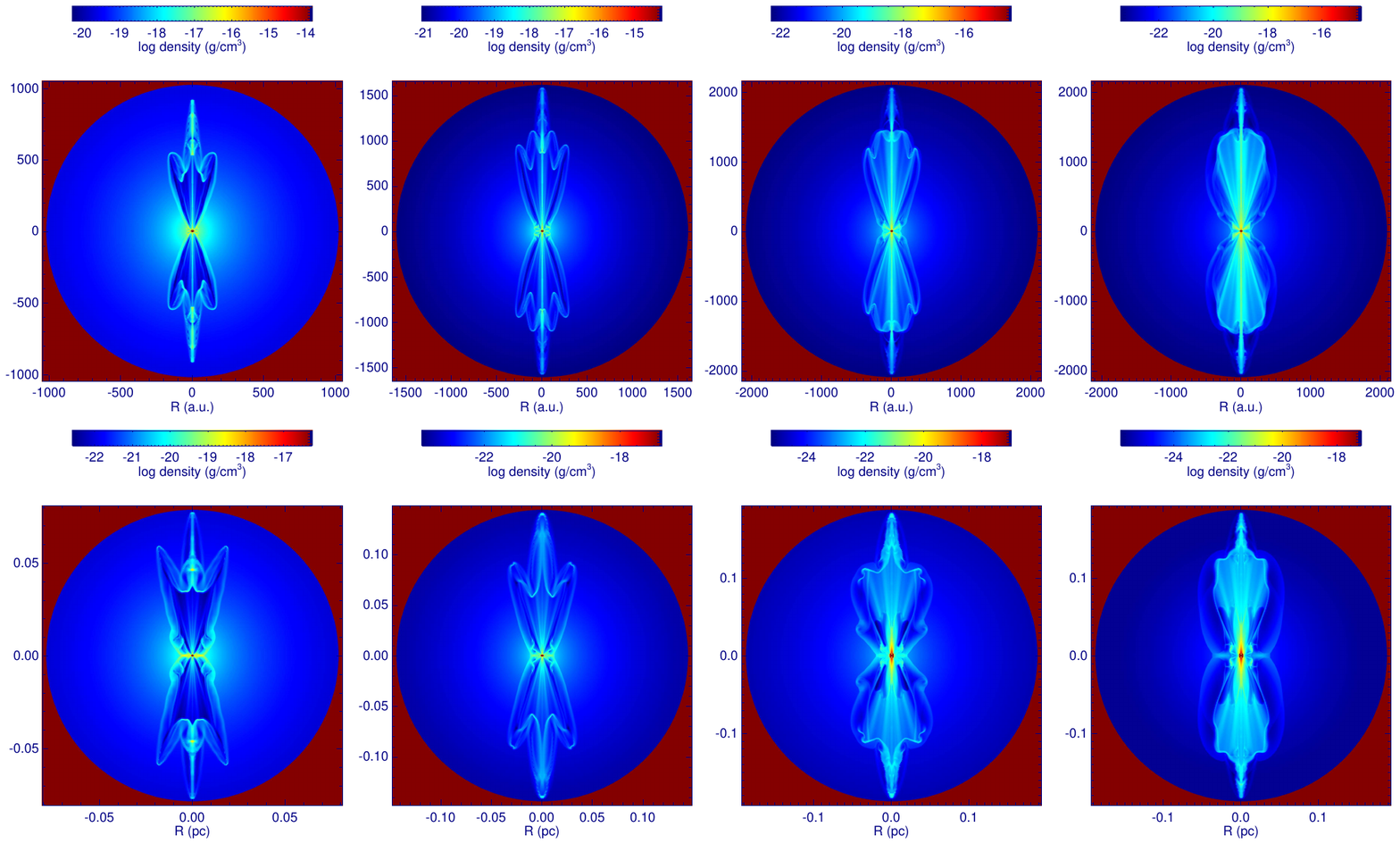}
\caption{Gas density snapshots at 50 yr (top) and 1,000 yr (bottom) of Models C6, C7, C8, and C9. }
\label{f12}
\end{figure}

\begin{figure}
\epsscale{1.25}
\plotone{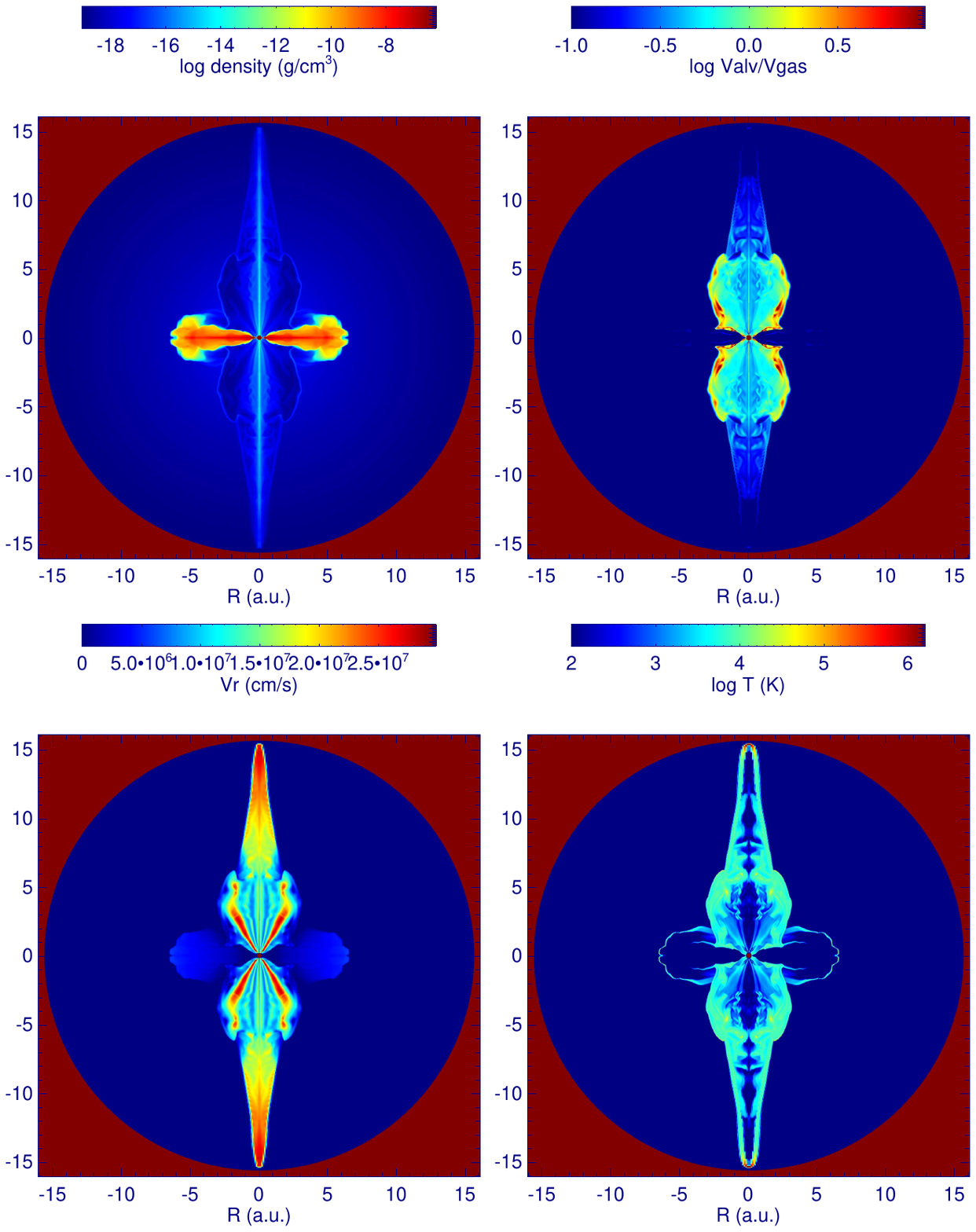}
\caption{Gas density, ratio of the Alfv\'en to the gas velocity, radial velocity, and temperature of Model D9 at 276.7 days after the  CE event.}
\label{f13}
\end{figure}

\begin{figure}
\epsscale{1.25}
\plotone{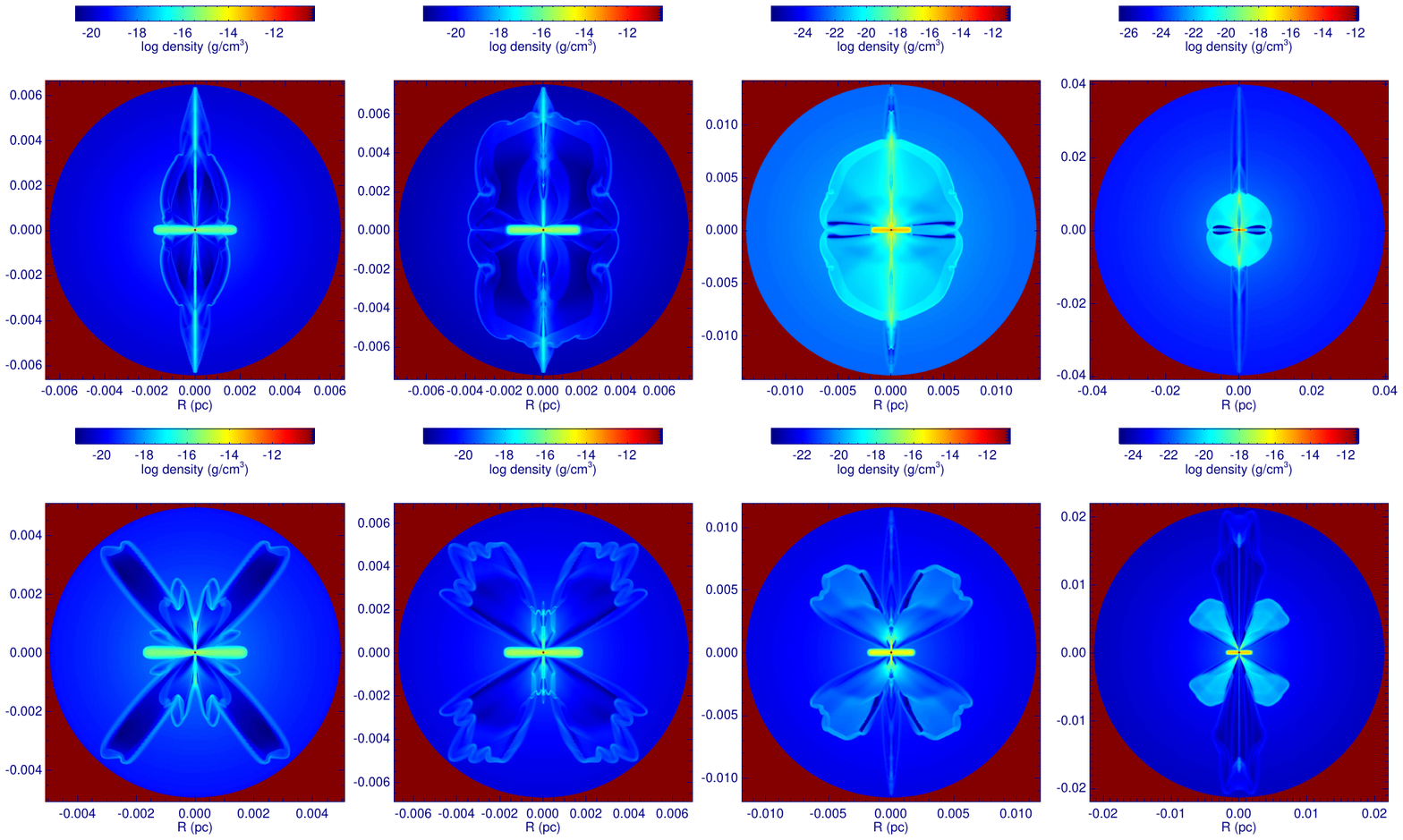}
\caption{Gas density snapshots of Models E6, E7, E8, E9 (top), and
F6, F7, F8, F9 (bottom) at 350 yr.}
\label{f14}
\end{figure}

\begin{figure}
\epsscale{1.30}
\plotone{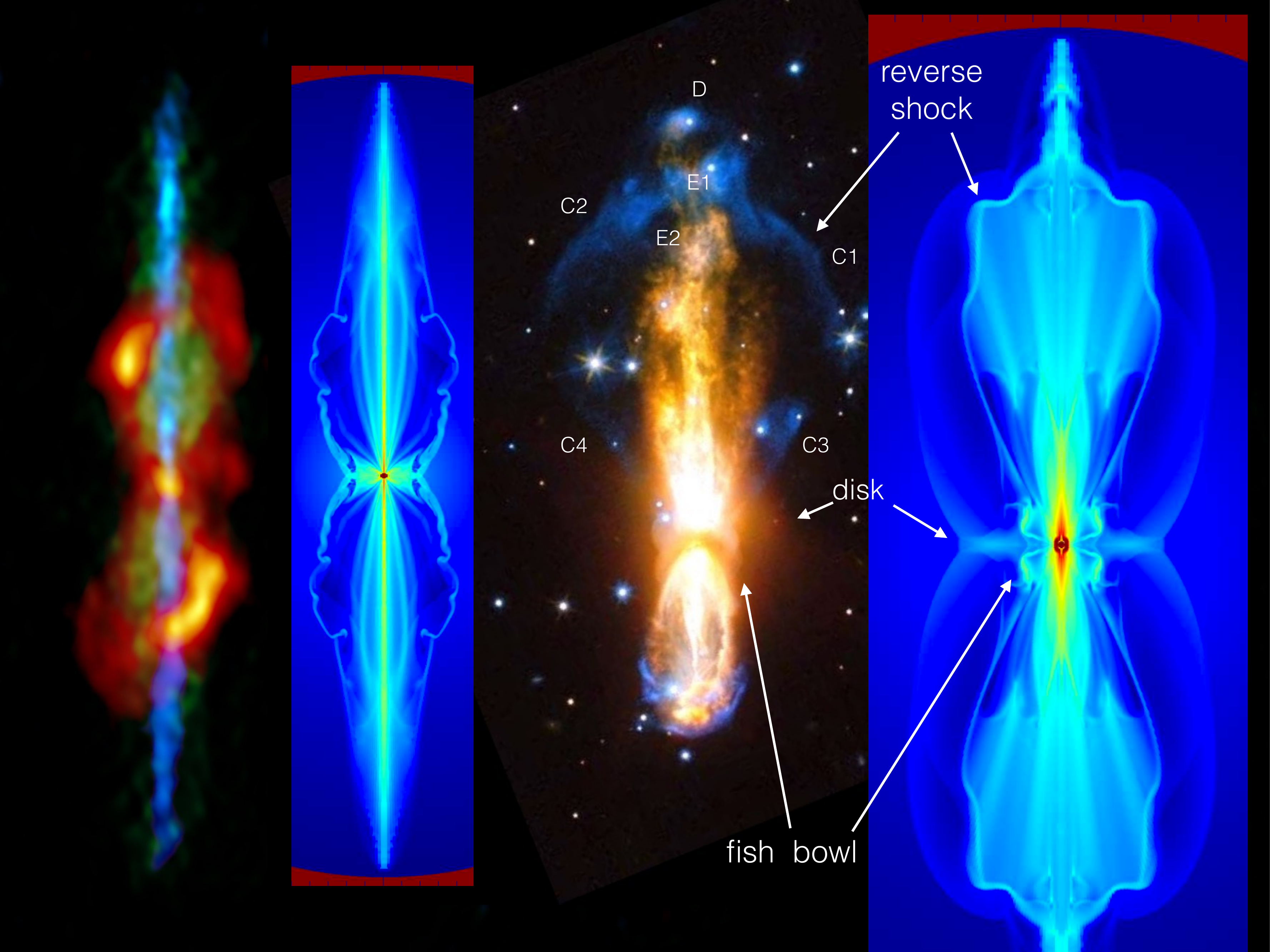}
\caption{The PPNe W43A ($\sim 60$ yr) with ALMA (left) and  OH231.8+4.2 ($\sim 870 $ yr)   with HST (right) in comparison with models B8 (50 yr) and C9 (1,000 yr)  respectively. W43A and OH231.8+4.2 are both 
inclined with respect to the plane of the sky by $i=35^{\circ}$.  Credits:  ALMA (ESO/NAOJ/NRAO), Tafoya et al.; NASA/ESA Hubble Space Telescope}
\label{f15}
\end{figure}

\begin{figure}
\epsscale{1.25}
\plotone{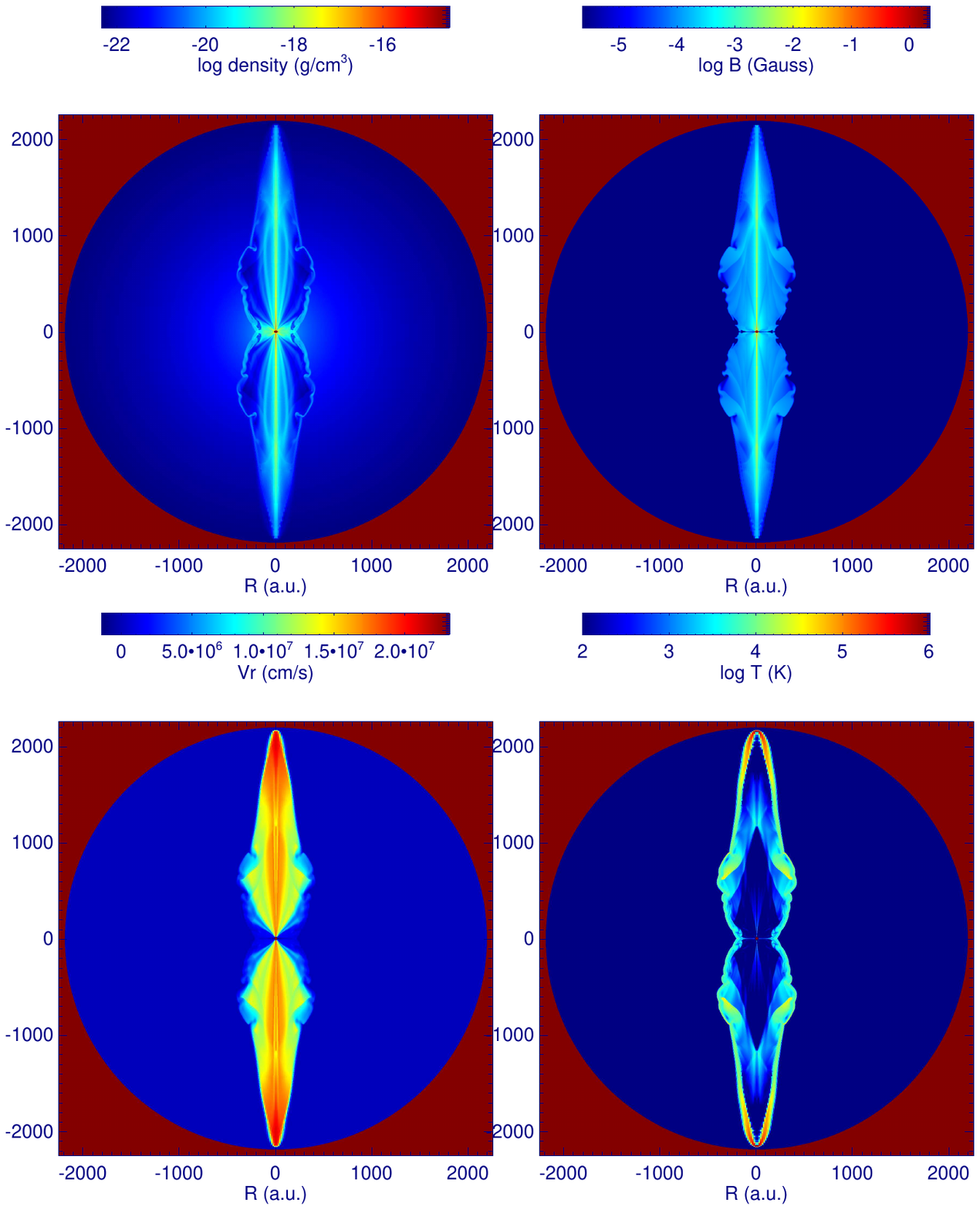}
\caption{Gas density, toroidal magnetic field, radial velocity, and temperature of Model B8 at 50 years after the  CE event.}
\label{f16}
\end{figure}

\begin{figure}
\epsscale{1.25}
\plotone{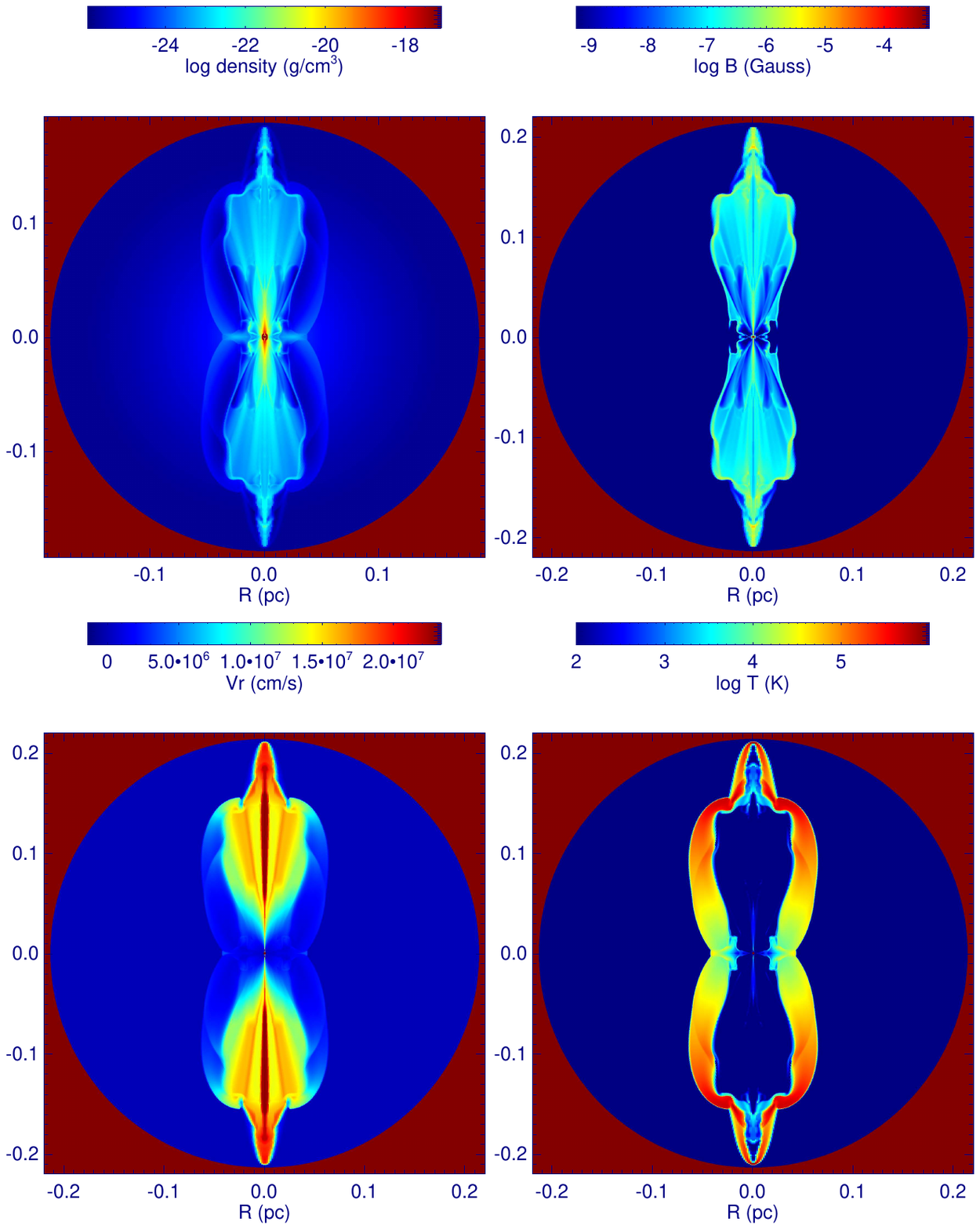}
\caption{Gas density, toroidal magnetic field, radial velocity, and temperature of Model C9 at 1,000 years after the  CE event.}
\label{f17}
\end{figure}

\end{document}